\documentclass[aps,prb,reprint,superscriptaddress,amsmath,amssymb]{revtex4-2}
\usepackage{bm,graphicx}
\newcommand{\dd}{\mathrm d}
\newcommand{\ii}{\mathrm i}

\begin{document}

\title{Classical weak antilocalization in electron hydrodynamics}

\author{Alberto Cortijo}
\email{alberto.cortijo@csic.es}
\affiliation{Instituto de Ciencia de Materiales de Madrid (ICMM), Consejo Superior de Investigaciones Cient\'{\i}ficas (CSIC), Sor Juana In\'es de la Cruz 3, 28049 Madrid, Spain}

\date{\today}

\begin{abstract}
Weak localization is strongly suppressed in electron hydrodynamics, where frequent electron--electron collisions rapidly dephase the Cooperon. Nevertheless, disorder can still produce a logarithmic correction to the bare hydrodynamic resistivity. We show that spatial fluctuations of the momentum-relaxation rate generate a classical memory kernel through the diffusive return of transverse momentum. In two dimensions, this return is logarithmic, as in weak localization, but involves a single shear propagator rather than quantum interference. Conservation laws and positivity of dissipation fix its antilocalizing sign: the return reduces momentum relaxation and hence the bulk resistivity.
Unlike quantum weak localization, the logarithm is cut off by momentum relaxation and has a viscosity-dependent amplitude. For temperature-independent impurity friction, the resistivity correction scales as $-T^2\ln(T_\xi/T)$ in a Fermi liquid and $-T\ln(T_\xi/T)$ in a Planckian metal. In channels, the return competes with viscosity renormalization, and a width scan separates the two effects.
\end{abstract}

\maketitle

\begin{table}[b]
\caption{\label{tab:compare}Two returns: weak (anti)localization in a diffusive metal versus the classical
friction return in an electron fluid. $\tau_{\rm mc}$ is the momentum-conserving
collision time, $\ell$ the mean free path; $\alpha=1$ (orthogonal) or $-\frac12$ (strong spin--orbit).}
\begin{ruledtabular}
\begin{tabular}{lll}
 & diffusive metal & electron fluid \\ \hline
regime & $\tau_{\rm mr}\ll\tau_{\rm mc}$ & $\tau_{\rm mc}\ll\tau_{\rm mr}$\\
transport & $\sigma=e^2N_FD$ & $\rho\simeq12\,m\nu/ne^2W^2$\\
soft mode & diffusion, $D$ & shear, $\nu=v^2/4\Gamma_2$\\
diagram & crossed ladder & single shear line\\
return & interference ($\hbar$) & classical\\
IR cutoff & $\tau_\varphi$ & $\tau_{\rm mr}=1/\gamma$\\
coupling & $1/\pi k_F\ell$ & $\sigma_1^2\xi^2/4D_\nu^2$\\
correction & $\frac{\delta\sigma}{\sigma}=-\frac{\alpha}{\pi k_F\ell}\ln\frac{\tau_\varphi}{\tau}$ & $\frac{\delta\gamma}{\gamma}\simeq-\frac{\sigma_1^2\xi^2}{4D_\nu^2}\ln\frac{1}{\gamma\tau_\xi}$\\
sign set by & phase of the pair & Wiener bound\\
sign & WL or WAL & anti-localizing only\\
\end{tabular}
\end{ruledtabular}
\end{table}

\section{Introduction}\label{sec:intro}

Quantum coherence modifies transport in disordered conductors through the
interference of time-reversed electronic paths~\cite{GLK1979,LeeRamakrishnan1985,Bergmann1984}.
In two dimensions, the probability for a diffusing electron to return to its
starting point decays with time as $P(t)=1/(4\pi Dt)$. Interference between a closed path
and its time reverse then produces a logarithmic correction to the conductivity,
$\delta\sigma=-(2e^2D/\pi\hbar)\int_\tau^{\tau_\varphi}\dd t\,P(t)
=-(e^2/2\pi^2\hbar)\ln(\tau_\varphi/\tau)$.
The logarithm reflects the marginal return probability in two dimensions and is
cut off by the dephasing time $\tau_\varphi$~\cite{AAK1982}. Its sign depends on
the symmetry class: ordinary weak localization enhances backscattering, whereas
strong spin--orbit coupling produces weak antilocalization~\cite{HLN1980,Bergmann1984}.
Hershfield and Ambegaokar (HA) recast this interference correction as a memory
term in the Boltzmann equation, with a kernel proportional to $P(t)$~\cite{HA1986}.

In electron hydrodynamics, however, this quantum return is strongly suppressed.
When momentum-conserving electron--electron collisions dominate, the electronic
Cooperon is expected to be dephased before impurity scattering can close a loop:
parametrically, $\tau_\varphi/\tau_{\rm mr}\lesssim\gamma/\Gamma_2\ll1$ in the
hydrodynamic window~\cite{NZA2002,LGL2019} (Appendix~\ref{app:cooperon}). Transport is then governed by collective viscous
flow rather than single-particle diffusion, with the Gurzhi effect as a
characteristic signature~\cite{Gurzhi1963,Gurzhi1968,Molenkamp1994,deJong1995,Narozhny2017,LucasFong2018,FritzScaffidi2024}.
Nevertheless, static disorder remains responsible for the finite momentum
relaxation and bulk resistivity. This raises the question that motivates our
work: \emph{does disorder merely set the bare hydrodynamic relaxation rates, or
can it generate an additional logarithmic correction to the resistivity even
when the Cooperon is dephased?} Quantum interference of hydrodynamic modes has
been investigated in a dirty marginal Fermi liquid~\cite{WuLiaoFoster2022};
the mechanism considered here instead belongs to classical viscous dynamics.

The distinction between the hydrodynamic relaxation rates is essential. The
long-wavelength modes comprise density ($m=0$), momentum ($m=\pm1$) and the
traceless stress ($m=\pm2$), labeled by angular harmonics on the Fermi circle
or, more generally, by scalar, vector and spin-two sectors. Electron--electron
collisions conserve density and momentum but relax the stress at a rate
$\Gamma_2$, which sets the viscosity $\nu=v^2/4\Gamma_2$. This is a coupling
of the clean fluid. By contrast, $\gamma=\tau_{\rm mr}^{-1}\ll\Gamma_2$
parametrizes the explicit breaking of translation invariance and is fixed by
the sample. Momentum is thus a weakly broken conservation law, the setting of
quasi-hydrodynamics~\cite{HLS2018,GLP2019}. Disorder may renormalize a
symmetry-breaking parameter differently from a coupling present in the clean
theory.

A logarithmic return requires two ingredients: a soft mode whose two-dimensional
return probability decays as $1/t$, and a disorder vertex that couples to it
without gradients. In an ordinary diffusive metal the conserved density is a
classical soft mode, but its exact continuity equation forces disorder to enter
through gradients, removing the infrared logarithm. The mode that has both
ingredients is the Cooperon, the propagator of a closed path together with its time
reverse. It is soft because, with time reversal, the two amplitudes accumulate the
same phase, and it is quantum: its weight carries $1/k_F\ell\propto\hbar$, and the
dephasing time cuts it off. An electron fluid has a different classical
candidate: the approximately conserved transverse momentum, with retarded
propagator $g_T(s,q)=[s+\gamma+\nu q^2]^{-1}$. It undergoes shear diffusion
and has a return probability $1/(4\pi\nu t)$, cut off at long times by
$\tau_{\rm mr}=1/\gamma$. A spatially varying friction couples directly to
this mode without gradients. Consequently, the return generates a logarithmic
memory kernel of the HA form, but it contains \emph{one retarded shear
propagator}, not a Cooperon or interference between time-reversed paths. The
overdamped shear mode carries no phase, so this return is classical: weak breaking of
a conservation law does in the fluid what time reversal does in the metal.

Conservation laws also constrain which disorder can generate this effect.
Number conservation requires the disorder contribution to the continuity
equation to be a divergence. Momentum conservation is broken by $\gamma$, so
within the local isotropic dissipative model considered here, the
momentum-relaxing perturbation is a random friction
$\gamma(\bm x)=\gamma[1+w_1(\bm x)]$. A second perturbation, a spatially
varying stress-relaxation rate $\Gamma_2(\bm x)=\Gamma_2[1+w_2(\bm x)]$,
conserves momentum. To all orders in the disorder, each perturbation corrects only
its own relaxation rate. The two also act differently: the friction couples to the
slow transverse momentum without gradients, so its correction is marginal in two
dimensions, whereas stress disorder reaches the momentum only through two gradients,
and its correction is irrelevant. Moreover,
momentum balance and the positivity of dissipation impose the Wiener bounds
$\langle\gamma(\bm x)^{-1}\rangle^{-1}\le\gamma_R
\le\langle\gamma(\bm x)\rangle=\gamma$. The fluid can rearrange its flow to
avoid high-friction regions, dissipating less than a uniform flow. The
correction thus \emph{lowers} the effective momentum-relaxation rate and the
bulk resistivity: its sign is necessarily antilocalizing, unlike the
symmetry-class-dependent sign of quantum interference.

At Born order, the friction return gives
$\gamma_R/\gamma=1-(\sigma_1^2/2)f(a_0)$, where
$a_0=\gamma\xi^2/2\nu$, $\xi$ is the disorder correlation length, and
$f(a)=ae^aE_1(a)$. In the hydrodynamic window $a_0\ll1$, this yields the
nonanalytic correction $\delta\gamma/\gamma\simeq-(\sigma_1^2/2)\,a_0\ln(1/a_0)$. The disorder vertex is proportional
to $\gamma$, and the same rate cuts off the shear propagator, so
$\delta\gamma/\gamma\propto\gamma\ln(1/\gamma)\to0$ as translation invariance
is restored. The distinction has an instructive parallel in chiral perturbation
theory: pion loops produce a mass-dependent logarithm in an explicitly
symmetry-breaking parameter, while couplings of the symmetric theory have a
different renormalization~\cite{GMOR1968,GasserLeutwyler1984}.

Stress disorder, on the other hand, renormalizes the viscosity, rather than
the uniform momentum-relaxation rate. Its leading correction survives when
$\gamma\to0$, while its temperature-dependent contribution comes from the same
shear return. For temperature-independent impurity friction and a
stress-relaxation rate proportional to $T^2$ in a Fermi liquid or $T$ in a
Planckian fluid, $a_0=(T/T_\xi)^p$ with $p=2$ or $1$, and the classical friction
correction scales respectively as
$-T^2\ln(T_\xi/T)$ and $-T\ln(T_\xi/T)$. In contrast with the quantum
weak-localization logarithm, its prefactor decreases as the fluid becomes more
viscous. Logarithmic refinements of $\Gamma_2(T)$ modify the map between
temperature and $a_0$; tomographic regimes with additional long-lived angular
harmonics lie beyond our Navier--Stokes truncation.

The competing corrections become distinguishable in a channel. For widths
$W\ll D_\nu=(\nu/\gamma)^{1/2}$, viscous dissipation dominates and the walls
cut off the long-distance return. For $W\gg D_\nu$, the resistivity approaches
the bulk value set by the renormalized friction. A family of channels of
different widths therefore separates the viscosity and friction corrections,
and isolates the nonanalytic temperature dependence. We also examine density
inhomogeneities, whose additional viscous dissipation~\cite{AKS2011} must be
distinguished from the classical return. Table~\ref{tab:compare} summarizes
the contrast with weak (anti)localization in a diffusive metal.

We work in two dimensions, within linear classical kinetic theory truncated
at the stress harmonic, in the regime $\ell_\nu\ll\xi\ll D_\nu$. The results
are given at Born order and checked through fourth order in disorder strength;
the structural constraints require only the conservation laws and the
hydrodynamic hierarchy, not quasiparticles. Section~\ref{sec:model} formulates
the model and the conservation-law bounds. Section~\ref{sec:return} derives the
friction return kernel and its renormalization of momentum relaxation.
Section~\ref{sec:stress} treats the stress sector, while
Sec.~\ref{sec:chanresp} develops the channel response, temperature dependence
and width-scan protocol. Section~\ref{sec:outlook} discusses other classical
memory effects and remaining limitations. Detailed derivations and checks are
in the appendices. We use $\hbar=k_B=1$.

\section{Model}\label{sec:model}

We linearize the distribution about equilibrium near the Fermi circle,
$f=f_0-(\partial f_0/\partial\varepsilon)\sum_m a_m(\bm x,t)\,e^{\ii m\theta}$, with $\theta$ the
direction of the momentum, and keep the conserved density ($m=0$), the momentum ($m=\pm1$) and the first
non-conserved harmonic, the traceless stress ($m=\pm2$):
\begin{equation}
\partial_t a_m+\tfrac v2\left(\partial_-a_{m-1}+\partial_+a_{m+1}\right)=-\gamma_m a_m ,
\label{eq:hier}
\end{equation}
with $\partial_\pm=\partial_x\pm\ii\partial_y$, $\gamma_0=0$,
$\gamma_{\pm1}=\gamma$, $\gamma_{\pm2}=\Gamma_2$, and $a_{\pm3}=0$. Eliminating the
stress row gives the transverse momentum (shear) propagator
\begin{equation}
g_T(s,\bm q)=\frac{1}{s+\gamma+\nu(s)q^2},\qquad \nu(s)=\frac{v^2/4}{s+\Gamma_2},
\label{eq:gT}
\end{equation}
so $\nu=v^2/4\Gamma_2$ and $D_\nu^2=\nu/\gamma$. We work in the hydrodynamic
window $\ell_\nu\ll\xi\ll D_\nu$, where $\xi$ is the disorder correlation length
and $\ell_\nu$ the nonlocality length of the viscosity; the truncation error is
$O(\ell_\nu^2/\xi^2)$ (Appendix~\ref{app:window}). The bulk corrections computed below depend on the relaxation
rates only through $a_0=\xi^2/2D_\nu^2$, which is small inside this window.

We add a local vertex $V(\bm x)$ to Eq.~(\ref{eq:hier}) and impose reality of
the fields, number conservation, Onsager reciprocity, positivity of the
dissipative part, and statistical isotropy. For non-magnetic scalar disorder these
conditions leave exactly two kinds of disorder (Appendix~\ref{app:admissible}):
\begin{equation}
V(\bm x)=\gamma\,w_1(\bm x)\,P_1+\Gamma_2\,w_2(\bm x)\,P_2 ,\qquad 1+w_a\ge 0 ,
\label{eq:V}
\end{equation}
where $P_{1,2}$ project on $m=\pm1,\pm2$. The first term is a random friction.
Positivity bounds it by $\gamma$, so it is slaved to the symmetry-breaking
parameter and vanishes as $\gamma\to0$. The second is a random stress-relaxation rate,
equivalently a random viscosity $\nu(\bm x)=\nu/(1+w_2)$. Couplings that would relax
momentum even at $\gamma=0$ are independent sources of breaking and are
excluded; the most relevant one, a density landscape, is added separately below. We take $w_{1,2}$ independent and, at the orders we compute, Gaussian; the
Born results depend only on the covariance, which a strictly positive ensemble reproduces, so the Gaussian stands for a
positive friction and stress-relaxation rate (Appendices~\ref{app:admissible} and \ref{app:window}). The covariance is
$\langle w_a(\bm x)w_a(\bm x')\rangle=\sigma_a^2e^{-|\bm x-\bm x'|^2/2\xi^2}$, a smooth correlator of
range $\xi$, with Fourier transform $C_a(k)=2\pi\sigma_a^2\xi^2e^{-k^2\xi^2/2}$. Only its integral, $C_a(0)$,
enters the coefficient of the logarithm; the shape enters only through a length under the logarithm, the mean
log-distance of the correlator (Appendix~\ref{app:born}). A density landscape, $n=\bar n(1+w)$, is a third and independent
amplitude $\sigma_n$: at second order it does not mix with $w_1$, and it adds the viscous dissipation of the flow
forced around the inhomogeneities, a relative increase $\sigma_n^2/2a_0$ of the bulk resistivity, the viscous term of Andreev,
Kivelson and Spivak~\cite{AKS2011} (Appendix~\ref{app:aks}). Their thermal term needs the energy sector and is outside the model.

Two exact statements follow from the Ward identity (Appendix~\ref{app:admissible}). Every term of the
self-energy begins and ends on a disorder vertex, so to all orders and for any
statistics each disorder acts only on its own harmonic: at $q=0$ stress disorder
leaves $\gamma$ exactly unchanged and friction disorder leaves $\Gamma_2$ exactly
unchanged. The
integrated momentum balance and positivity give the Wiener bounds
$\langle\gamma(\bm x)^{-1}\rangle^{-1}\le\gamma_R\le\langle\gamma(\bm x)\rangle=\gamma$,
the same bounds that hold for the conductivity of a composite~\cite{Torquato2002}. They have a
simple meaning. The upper bound, the arithmetic mean, is the friction of a flow that stays uniform and
crosses every patch, high-friction or not, at the same velocity. The actual flow dissipates less,
because it bends around the high-friction patches, and of all flows carrying the same current the one
realized is the one that dissipates least. The lower bound, the harmonic mean, is the friction the
fluid would reach if it could redistribute freely, each patch carrying a velocity inversely proportional
to its friction; incompressibility and viscosity keep it away from that limit. The upper bound fixes
the sign: no static disorder of this kind can increase momentum relaxation.

\section{Return kernel}\label{sec:return}

\begin{figure}[t]
\includegraphics[width=\columnwidth]{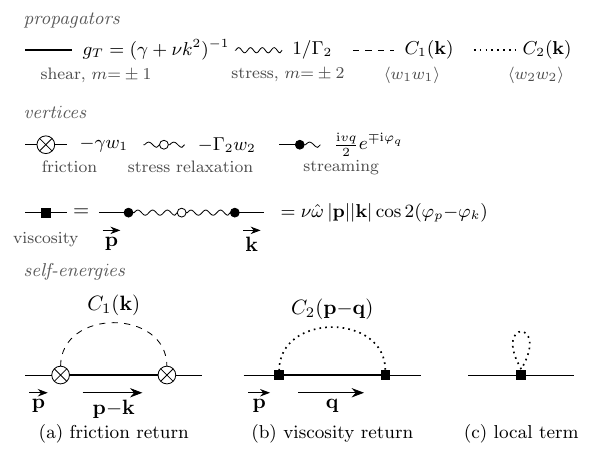}
\caption{\label{fig:diagrams}Diagrammatic dictionary. Top: propagators, the transverse momentum (shear) line
$g_T$, the stress line $1/\Gamma_2$, and the friction and stress-disorder correlators $C_1$ and $C_2$. Middle:
vertices, friction $-\gamma w_1$, stress relaxation $-\Gamma_2w_2$ and streaming $(\ii vq/2)e^{\mp\ii\varphi_q}$; the
viscosity vertex $\blacksquare$ is the stress row integrated out (streaming into $m=\pm2$, one stress-relaxation insertion,
streaming back), with $\hat\omega$ the Fourier component of $\omega=1/(1+w_2)-1$. Bottom: Born self-energies of the
transverse momentum line, (a) the friction return, Eq.~(\ref{eq:Sigma11}), and (b) the viscosity return and (c) the
local (Voigt) term, both of order $p^2$.}
\end{figure}

In $m$ space the disorder-averaged self-energy is diagonal. At Born order
the friction self-energy of the momentum harmonic is diagram (a) of Fig.~\ref{fig:diagrams}:
\begin{equation}
\Sigma_{11}(s)=\gamma^2\!\int\!\frac{\dd^2k}{(2\pi)^2}\,C_1(\bm k)\,g_T(s,\bm k)\,[T_{\bm k}]_{xx},
\label{eq:Sigma11}
\end{equation}
where $[T_{\bm k}]_{ij}=\delta_{ij}-\hat k_i\hat k_j$ is the transverse
projector. Only the transverse part survives at DC, because continuity freezes
longitudinal momentum. Its angular average $\int_0^{2\pi}\frac{\dd\varphi}{2\pi}\sin^2\varphi=\frac12$
is the transverse weight. Transformed to real time, this self-energy adds a memory
term to the kinetic equation of the momentum harmonic. At zero wave vector the
momentum equation becomes
$\partial_t\bm u=-\gamma\bm u+\int_{-\infty}^t\dd t'\,K(t-t')\,\bm u(t')+\dots$,
with
\begin{equation}
K(t)=\frac{\gamma^2}{2}\,C_1(0)\,P_\nu(t)\,e^{-\gamma t},\qquad
P_\nu(t)=\frac{1}{4\pi\nu\,(t+\tau_\xi)},
\label{eq:K}
\end{equation}
and $\tau_\xi=\xi^2/2\nu$. $P_\nu$ is the return probability of shear diffusion,
smeared over $\xi$: momentum lost to a friction patch at time $t'$ partly returns
at a later time $t$, carried by the shear mode. This is the HA
structure~\cite{HA1986}, with $\nu$ in place of $D$ and momentum relaxation in
place of dephasing.

Integrating Eq.~(\ref{eq:K}) over time gives the correction to $\gamma$. In a
steady flow the memory term acts as an extra, negative friction, so the
momentum-relaxation rate seen in DC transport is
$\gamma_R=\gamma-\int_0^\infty\dd t\,K(t)$:
\begin{equation}
\frac{\gamma_R}{\gamma}=1-\frac{\sigma_1^2}{2}f(a_0)\simeq1-\frac{\sigma_1^2\xi^2}{4D_\nu^2}\Big[\ln\frac{1}{\gamma\tau_\xi}-\gamma_E\Big],
\label{eq:gammaR}
\end{equation}
where $a_0=\xi^2/2D_\nu^2=\gamma\tau_\xi$ and
\begin{equation}
\begin{gathered}
f(a)=\int_0^\infty\dd u\,e^{-u}\,\frac{a}{a+u}=a\,e^{a}E_1(a),\\
E_1(a)=\int_a^\infty\frac{\dd t}{t}\,e^{-t},
\end{gathered}
\label{eq:fdef}
\end{equation}
with $E_1$ the exponential integral. The function $f$ rises monotonically from $f\simeq a[\ln(1/a)-\gamma_E]$ at
$a\ll1$, with $\gamma_E\simeq0.577$ Euler's constant, to $f\to1$ at $a\gg1$. Here $\gamma$ is the
bare relaxation rate, the sample average $\langle\gamma(\bm x)\rangle$ that enters the
kinetic equation and sets both the mass of the shear line and $a_0$;
Eq.~(\ref{eq:gammaR}) is explicit in it. $\gamma_R$ is the renormalized rate, the one
a bulk resistivity measurement sees. The
logarithm is the time integral of $P_\nu\propto1/t$, marginal in two dimensions.
The sign is the anti-localizing one allowed by the Wiener bound. Physically, the
fluid rearranges to avoid high-friction patches, and the returning shear
perturbation is that rearrangement seen in time. The coupling
$\sigma_1^2\xi^2/4D_\nu^2$ is tied to the infrared length itself and vanishes as
$\gamma\to0$. Placing the HA kernel~\cite{HA1986} on the same footing confirms
the dictionary of Table~\ref{tab:compare}, with HA's prefactor fixed
independently by the Gor'kov--Larkin--Khmel'nitskii result~\cite{GLK1979} (Appendix~\ref{app:born}).

The same calculation has a transparent hydrodynamic reading. At DC, eliminating
the stress row turns Eqs.~(\ref{eq:hier}) and (\ref{eq:V}) exactly into a random
Brinkman problem for the velocity field,
\begin{equation}
\gamma(\bm x)\,u_i+\partial_iP-\partial_j\big[\nu(\bm x)\,S_{ij}\big]=F_i,\qquad \partial_iu_i=0,
\label{eq:brinkman}
\end{equation}
with $\gamma(\bm x)=\gamma(1+w_1)$, $\nu(\bm x)=\nu/(1+w_2)$, $S_{ij}=\partial_iu_j+\partial_ju_i$ the strain rate, $P$ the
pressure and $\bm F$ the driving force, both per unit mass. The viscous stress is $\nu(\bm x)S_{ij}$; the symmetric
form matters because $\nu$ varies in space, and for a uniform viscosity the viscous term reduces to $\nu\nabla^2u_i$. Brinkman
introduced this equation for slow flow through porous media, where the friction
term is the drag of the solid matrix and the viscous term the shear of the fluid
between its grains~\cite{Brinkman1949,KimRussel1985}. With uniform coefficients it
is the familiar description of viscous electron flow with momentum relaxation,
used for instance for current flow around obstacles and
constrictions~\cite{Pawlowski2026}; here the coefficients are random. Two limits
anchor the result. When the disorder varies on scales much longer than the Gurzhi
length, $a_0\to\infty$, viscosity drops out and Eq.~(\ref{eq:brinkman}) reduces to
Darcy's law, $\gamma(\bm x)\bm u=\bm F-\nabla P$: a divergence-free current in a
medium of random local conductivity $1/\gamma(\bm x)$. In two dimensions Dykhne
showed, by a duality between current and field, that the effective conductivity of
such a medium is exactly the geometric mean of the local one when the logarithm of
the conductivity is symmetrically distributed~\cite{Dykhne1971,Matheron1967}. At
second order this gives $\gamma_R/\gamma=1-\sigma_1^2/2$, the $a_0\to\infty$ limit of
Eq.~(\ref{eq:gammaR}), since $f(\infty)=1$ (Appendix~\ref{app:born}). In the opposite limit,
$a_0\to0$, the logarithm in Eq.~(\ref{eq:gammaR}) is the two-dimensional Stokes
logarithm, cut off by the Gurzhi length $D_\nu$ as for a no-slip obstacle in an
electron fluid~\cite{Lucas2017Stokes}. Exact numerical solutions of
Eq.~(\ref{eq:brinkman}) confirm the Born results, and the fourth-order corrections
need no resummation (Appendix~\ref{app:beyond}). These connections give the calculation independent
anchors. What the electron fluid adds is the structure behind them: the friction
contrast is bounded by $\gamma$, so the return vanishes with the
symmetry-breaking parameter instead of being fixed by geometry, and
$a_0=\gamma\tau_\xi$ turns the infrared cutoff into a temperature. The effect is
distinct from the classical memory effects of the two-dimensional electron gas,
which come from ballistic returns of single electrons to the same scatterer,
mainly in a magnetic field~\cite{Mirlin2001,Raskulov2026}, and from the
long-time tails of fluctuating hydrodynamics, which need thermal noise and
nonlinear advection~\cite{FNS1977,KMR2011} and are absent from the linear
theory used here. The diagrammatic structure also differs from weak localization.
In the metal the return is carried by a composite mode: the Cooperon, the ladder
of a path and its time reverse, is gapless because time reversal maps it onto the
diffuson, whose pole is protected by number conservation, and only dephasing cuts
it off. In the fluid the return is carried by a single shear line, and no
conservation law protects a pair of them. A conserving treatment (self-consistent
Born plus ladder) places the would-be pair pole, $1/(2\gamma+\nu Q^2/2-\ii\Omega)$,
exactly at the edge of the two-line continuum of the shear mode, so it is not a
bound state; exact spectra of the random Brinkman operator show a continuous edge
with a Lifshitz tail and no isolated level (Appendix~\ref{app:beyond}).

\section{Stress sector}\label{sec:stress}

Stress disorder enters the $m=\pm2$ row. Integrating out that row in each
realization replaces $\Gamma_2(1+w_2)$ by the local viscosity $\nu/(1+w_2)$. Its
first order in $w_2$ is the viscosity vertex $\blacksquare$ of
Fig.~\ref{fig:diagrams}, whose spin-two angular factor comes from the phases of
the two streaming steps; its second order at one point is the local
insertion~(c). Both vanish for uniform flow, which has no strain, so they
renormalize $\nu$ and not $\gamma$. At Born order
\begin{equation}
\frac{\delta\nu}{\nu}=\frac{\sigma_2^2}{2}\big[1+f(a_0)\big] .
\label{eq:dnu}
\end{equation}
The two parts reflect the status of $\Gamma_2$. The local part $\sigma_2^2/2$ does not depend on $\gamma$: $\Gamma_2$
is a coupling of the clean fixed point and is shifted already at leading order, by a constant that a measurement of
the viscosity absorbs. The $f(a_0)$ part is the stress fluctuation converting into shear momentum and returning. It
carries the temperature dependence of the correction and vanishes with $\gamma$, so that temperature dependence is
induced by the momentum relaxation. The viscosity rises in both parts. In real time the viscosity memory is
\begin{equation}
\delta\nu(t)=\nu\sigma_2^2\Big[\delta(t)-\frac{\tau_\xi}{2}\,\frac{e^{-\gamma t}}{(t+\tau_\xi)^2}\Big],
\label{eq:dnut}
\end{equation}
valid for $t\gg1/\Gamma_2$ and $t$ longer than the sound crossing time of $\xi$.
The two gradients of the stress vertex add one power of $1/(t+\tau_\xi)$ relative
to Eq.~(\ref{eq:K}), so the return tail is integrable at $\gamma=0$. At $\gamma=0$ the correction, $\sigma_2^2/2$, is fixed
independently by an exact duality of two-dimensional Stokes flow, the viscous
counterpart of Dykhne's result, which gives the geometric mean
$\nu\,e^{\sigma_2^2/2}$ (Appendix~\ref{app:beyond}).

\begin{figure*}[t]
\includegraphics[width=0.92\textwidth]{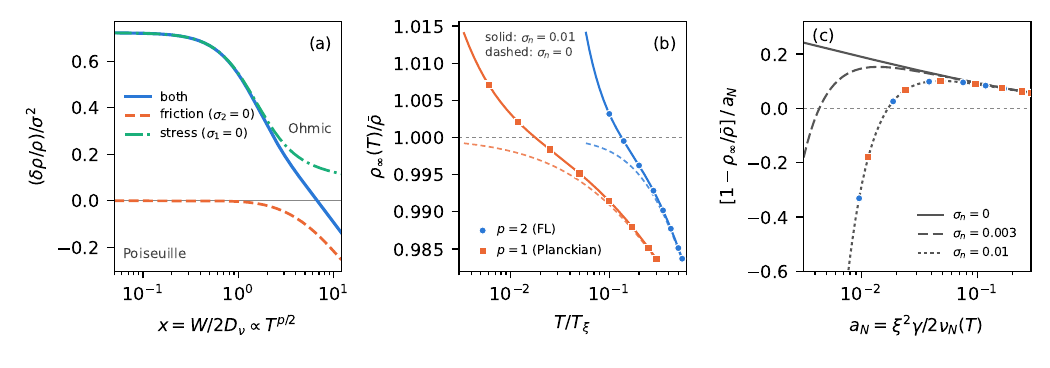}
\caption{\label{fig:widthscan}Channel response. (a) Born-order correction to the channel resistivity, Eq.~(\ref{eq:channel}), against $x=W/2D_\nu\propto T^{p/2}$, in units of $\sigma^2$ ($\sigma_1=\sigma_2=\sigma$), for $W=20\,\xi$ with the wall term included: friction only, stress only, and both. The viscosity correction dominates in the Poiseuille regime and the friction return in the Ohmic regime; the total changes sign in between. (b), (c) Width-scan protocol. Points: Born channel formula, Eq.~(\ref{eq:channel}) with the
density-landscape term, at widths $W=100$--$400\,\xi$ for $v\tau_{\rm mr}=1000\,\xi$ and $\sigma_1=\sigma_2=0.3$,
extrapolated to $\xi/W\to0$ with the uniform-channel shape; a theory illustration, not data. Fermi liquid ($p=2$, circles)
and Planckian fluid ($p=1$, squares), $a_0=(T/T_\xi)^p$. (b) Bulk intercept $\rho_\infty$ in units of
$\bar\rho=\rho_D(1+\sigma_n^2/2)$, with (solid) and without (dashed) a density landscape $\sigma_n=0.01$.
(c) $[1-\rho_\infty/\bar\rho]/a_N$ against $a_N$ from the narrow-channel slope, for $\sigma_n=0$, $0.003$ and $0.01$;
points for $\sigma_n=0.01$. Both fluids collapse on one curve.}
\end{figure*}

\section{Channel response}\label{sec:chanresp}

In a channel the two relaxation rates compete. Take a strip of width $W$ with no-slip walls, driven by a uniform
force. Consider first a uniform channel, in which disorder enters only through its average effect, the friction
$\gamma$, and the stress-relaxation rate is uniform too. The flow profile then results from a competition between
viscosity, which carries the no-slip condition from the walls into the fluid over the Gurzhi length $D_\nu$, and the
friction $\gamma$, which acts everywhere. When the
channel is narrower than $D_\nu$, viscosity wins: the profile is parabolic (Poiseuille), and the resistivity is the
viscous Gurzhi value $12m\nu/ne^2W^2$, which falls with temperature as $\nu$ does. When it is wider, friction wins: the
flow is uniform except in viscous boundary layers of width $D_\nu$ at the walls, and the resistivity approaches the
Ohmic Drude value $\rho_D=m\gamma/ne^2$, increased by the boundary layers to $\rho_D(1+2D_\nu/W)$. The full crossover is
$\rho_{\rm ch}=\rho_D/h(x)$, with $h=1-\tanh x/x$ and $x=W/2D_\nu$~\cite{Gurzhi1968,deJong1995}.

A small change of either relaxation rate changes $\rho$ in proportion to the share of the dissipation it controls:
\begin{equation}
\frac{\delta\rho}{\rho}=(1-\eta)\,\frac{\delta\gamma}{\gamma}+\eta\,\frac{\delta\nu}{\nu}+\Delta_w ,\qquad
\eta=\frac{\tanh x/x-{\rm sech}^2x}{2(1-\tanh x/x)} .
\label{eq:channel}
\end{equation}
Here $\eta=\partial\ln\rho/\partial\ln\nu$, and it equals the fraction of the dissipated power that is viscous. The
reason is the minimum principle met in Sec.~\ref{sec:model}: the realized flow dissipates least among all flows carrying
the same current, so a small change of $\nu$ or $\gamma$ changes the dissipation only through its explicit appearance,
not through the change of the profile, and each rate enters with its own share. The two weights add to one because
$\rho$ scales linearly when $\gamma$ and $\nu$ are rescaled together. In the Poiseuille regime $\eta\to1$, and friction
hardly matters; in the Ohmic regime $\eta\simeq1/2x=D_\nu/W$, because only the boundary layers dissipate viscously.
Narrow channels therefore see mainly the viscosity correction, $\delta\rho/\rho\simeq+\sigma_2^2/2+O(\xi/W)$, positive
because stress disorder raises the viscosity; wide channels see mainly the friction return,
$\delta\rho/\rho\simeq-(\sigma_1^2/2)f(a_0)$, negative because the return lowers the friction. The correction changes
sign in between [Fig.~\ref{fig:widthscan}(a)].

The last term in Eq.~(\ref{eq:channel}) is a wall effect of stress disorder. In the bulk the flow partly avoids a
patch of high viscosity: it shears less inside the patch and more around it, so the dissipation rises by less than
the local viscosity does. This rearrangement is what halves the local (Voigt) shift of the viscosity, $\sigma_2^2$, to
the $\sigma_2^2/2$ of Eq.~(\ref{eq:dnu}) at $\gamma=0$. Within a distance of order $\xi$ from a no-slip wall the flow
cannot rearrange as freely, because the wall pins the velocity; the viscosity fluctuations there are less relaxed, and
this thin layer dissipates more than the bulk. Extra dissipation confined to a thin layer at the wall acts on the rest
of the flow as if the no-slip plane had moved into the fluid, that is, as a Navier slip length of negative sign,
an anti-slip,
$\ell_s=-\frac{14}{15}\sqrt{2/\pi}\,\sigma_2^2\xi\simeq-0.74\,\sigma_2^2\xi$, a small fraction of $\xi$
(Appendix~\ref{app:channel}). The channel then behaves as if it were narrower by $2|\ell_s|$, which raises its
resistivity by $\Delta_w=-(2\ell_s/W)\tanh^2x/h(x)$: by $6|\ell_s|/W\simeq4.5\,\sigma_2^2\xi/W$ in the Poiseuille
regime and by $2|\ell_s|/W$ in the Ohmic one. This anti-slip does not depend on temperature, which distinguishes it from
the kinetic slip of a real boundary, a boundary property outside the present model.

Temperature enters through a single parameter. The bulk corrections depend on the bare relaxation rates only through
$a_0=\gamma\tau_\xi=2\gamma\Gamma_2\xi^2/v^2$, the ratio between the time $\tau_\xi$ that shear takes to diffuse across a
disorder correlation length and the momentum lifetime $1/\gamma$. With $\gamma$ fixed and $\Gamma_2\propto T^p$, the
fluid becomes less viscous as the temperature rises, $\tau_\xi$ grows, and $a_0=(T/T_\xi)^p$, where $T_\xi$ is the
temperature at which $D_\nu=\xi/\sqrt2$. The channel parameter grows in the same way, $x^2=(W/\xi)^2a_0/2\propto T^p$,
so heating moves a channel of fixed width from the narrow to the wide side of Fig.~\ref{fig:widthscan}(a). The window
$\ell_\nu\ll\xi\ll D_\nu$ bounds the temperature range from both sides: at low temperature the viscosity becomes nonlocal
on the scale $\xi$, and at high temperature the Gurzhi length falls below $\xi$ (Appendix~\ref{app:window}). Since only
the hydrodynamic modes enter, the same holds at a non-Fermi-liquid fixed point, with $m=0,\pm1,\pm2$ read as the scalar,
vector and spin-two sectors~\cite{HLS2018}. The two ends of Eq.~(\ref{eq:channel}) then have very different
temperature dependences.

Wide channels carry the signature. Their width-independent part, the bulk intercept $\rho_\infty$, is the renormalized
friction itself, $\rho_\infty/\rho_D=\gamma_R/\gamma$ without a density landscape. To fourth order in the disorder its
logarithms are
\begin{equation}
\frac{\rho_\infty}{\rho_D}=1-\frac{\sigma_1^2}{2}\Big(1-\frac{\sigma_2^2}{2}\Big)a_0\ln\frac{1}{a_0}
+\frac{\sigma_1^2\sigma_2^2}{4}a_0^2\ln^2\frac{1}{a_0}+\dots ,
\label{eq:wide}
\end{equation}
where the dots are non-logarithmic terms of order $a_0$ and terms of order $a_0^2\ln(1/a_0)$ (Appendix~\ref{app:beyond}).
By confinement, stress disorder enters only through the mixed fourth-order terms: it weakens the friction logarithm by
the factor $1-\sigma_2^2/2$ and adds a squared logarithm. At low temperature the leading term gives
$\delta\rho/\rho\simeq-(p\sigma_1^2/2)(T/T_\xi)^p\ln(T_\xi/T)$: $-T^2\ln(T_\xi/T)$ in a Fermi liquid ($p=2$) and
$-T\ln(T_\xi/T)$ for a Planckian stress-relaxation rate ($p=1$). This is the hydrodynamic analogue of the
weak-localization logarithm, with the anti-localizing sign fixed by the Wiener bound, but its temperature dependence is
the opposite: the weak-localization logarithm grows as the temperature falls because dephasing slows down, whereas this
one is cut off by $\gamma$, and its prefactor vanishes as the fluid becomes more viscous at low temperature. It cannot be
absorbed into the bare relaxation rates. A temperature-independent $\gamma$ and a viscosity $\nu\propto T^{-p}$ give wide
channels only a constant and the boundary-layer contribution $\propto T^{-p/2}$, so a $T^p\ln(1/T)$ term in $\rho(T)$ is a
fingerprint of the return (Appendix~\ref{app:channel}).

Narrow channels have no such term, because the walls cut off the return. Shear diffusion cannot spread farther than $W$,
so the return integral never reaches the infrared, and its logarithm becomes $\ln(W/\xi)$. This holds beyond
perturbation theory. In a strip of width $W$ every shear flow relaxes at least at the rate $\pi^2\nu_{\min}/W^2$ of the
slowest viscous mode across the channel, set by the smallest local viscosity, and friction only makes relaxation
faster. The walls therefore keep a gap in the relaxation spectrum, and with a gap the resistivity depends smoothly on
$\gamma$: it contains only integer powers of $x^2\propto T^p$, with coefficients carrying $\ln(W/\xi)$
(Appendix~\ref{app:channel}). \emph{The narrow-channel resistivity is analytic in $x^2$, i.e. in $1/\nu_N$, at any
order in the disorder} (order by order in $\sigma$ for Gaussian disorder): it contains no $\ln a_0$, and its temperature
dependence is that of $x^2\propto\gamma\Gamma_2$ alone. The fingerprint of the return is therefore a term non-analytic in
$a_0$, present in wide channels and absent in narrow ones; it appears as a $\ln T$ in $\rho(T)$ when $\Gamma_2$ is a pure
power of $T$, the case considered here. Disorder is in fact nearly invisible there. In the narrow-channel resistivity,
$\rho=\rho_P+\frac65\rho_D-\frac{3}{175}(\rho_D^2/\rho_P)(1+\Delta)+\dots$, the three terms are the viscous dissipation
of the Poiseuille profile, $\rho_P=12m\nu/ne^2W^2$, the friction acting on that profile, and the first correction to the
profile caused by friction; with $\nu\propto T^{-p}$ and $\gamma$ fixed they scale as $T^{-p}$, $T^0$ and $T^p$. To
order $\sigma^2$ and $\xi/W$, disorder only renormalizes the two relaxation rates: stress disorder raises $\nu$ by
$\sigma_2^2/2$, friction disorder enters only through its mean $\gamma$, and the wall slip is equivalent to a small
change of both. Then $\Delta=0$, and a single narrow channel cannot be told from a uniform one with different relaxation
rates (Appendix~\ref{app:channel}). The return enters only at order $(\xi/W)^2$, $\Delta\sim\sigma^2(\xi/W)^2\ln(W/\xi)$;
for $W=20\,\xi$ and $\sigma_{1,2}=0.3$, $\Delta\approx0.08$, but it multiplies a term of order $10^{-4}$--$10^{-3}$ of
$\rho$, so it is a test in principle only. What narrow channels do provide is a direct measurement of the viscosity.

This is what makes a scan over widths work. A single sample cannot tell the bare from the corrected relaxation rates;
a family of channels of different widths, cut from the same material, can, because each relaxation rate is measured
in the channels where it dominates. In narrow channels the leading Poiseuille term measures $\nu_N=\nu(1+\sigma_2^2/2)$, the
bare $\nu(T)$ times a temperature-independent factor, because the walls cut off the temperature-dependent return part
of the viscosity correction; the disorder slip enters only odd powers of $1/W$. Wide channels, extrapolated to
$\xi/W\to0$, give the bulk intercept
\begin{equation}
\begin{gathered}
\frac{\rho_\infty}{\rho_D}=1+\frac{\sigma_n^2}{2}+\frac{\sigma_n^2}{2a_0}-\frac{\sigma_1^2}{2}f(a_0),\\
a_0=\Big(1+\frac{\sigma_2^2}{2}\Big)a_N,\qquad a_N=\frac{\xi^2\gamma}{2\nu_N},
\end{gathered}
\label{eq:intercept}
\end{equation}
so the measured $\nu_N(T)$ fixes $a_0(T)$ up to one constant factor, and the landscape does not change it. The
intercept then has a known form as a function of the measured $\nu_N$: at leading logarithm it is a combination of $1$,
$\nu_N$, $\nu_N^{-1}\ln\nu_N$ and $\nu_N^{-1}$. These four functions of temperature are independent, so a fit returns
the four unknowns, $\rho_D$, $\xi^2(1+\sigma_2^2/2)$, $\sigma_1^2$ and $\sigma_n^2$, uniquely (Appendix~\ref{app:widthscan}).
Figures~\ref{fig:widthscan}(b) and \ref{fig:widthscan}(c) show the protocol on data computed from the theory. Plotted
against $a_N$, the intercepts of a Fermi liquid and of a Planckian fluid fall on one curve, because both depend on
temperature only through $a_0$: uniform relaxation rates give zero, the friction return a slope in $\ln a_N$, and the density
landscape a downturn $\propto-\sigma_n^2/a_N^2$ at small $a_N$, which crosses zero at $a_0f(a_0)=\sigma_n^2/\sigma_1^2$.
The two contributions have opposite signs, the landscape adding viscous dissipation and the return reducing friction,
and they dominate at opposite ends of the temperature range, which is what makes them separable. The figure uses
$\sigma_{1,2}=0.3$, where the fourth-order terms amount to about $11\%$ of the Born correction; they stay below $10\%$ for
$\sigma_2\lesssim0.28$ (Appendix~\ref{app:widthscan}).

\section{Outlook}\label{sec:outlook}

In the hydrodynamic regime the quantum return is lost. Electron--electron
collisions dephase the Cooperon long before impurity scattering can close a loop,
as has been known since the work of Altshuler, Aronov and
Khmelnitskii~\cite{AAK1982,NZA2002,LGL2019}. Returns as such are not lost. The
fluid has its own soft modes, the quasi-conserved transverse momentum among them,
and their return probability is classical and still marginal in two dimensions.
The friction return computed here is one example. The long-time tails of
fluctuating hydrodynamics are another: thermal fluctuations of the same shear mode,
coupled by advection, return and renormalize the viscosity with a two-dimensional
logarithm~\cite{FNS1977,KMR2011}, which in an electron fluid would also be cut off
by $\gamma$. The ballistic memory effects of the electron gas~\cite{Mirlin2001}
are a third, in the opposite regime. Corrections of the weak-localization type in an
electron fluid are therefore a question about classical returns, and the Ward
identities decide which relaxation rate each of them corrects.

Phonons, left out so far, make the momentum-relaxation rate temperature dependent. If their friction is uniform,
$\gamma(\bm x)=\gamma_{\rm ph}+\gamma_i[1+w_1(\bm x)]$, they enter in a simple way. The disorder vertex is the
impurity part alone, while the shear line decays at the total rate $\gamma_{\rm tot}=\gamma_{\rm ph}+\gamma_i$, so that
\begin{equation}
\gamma_R\simeq\gamma_{\rm tot}-\frac{\sigma_1^2\gamma_i^2\xi^2}{4\nu}\Big[\ln\frac{2\nu}{\gamma_{\rm tot}\xi^2}-\gamma_E\Big].
\end{equation}
The phonon friction adds to the Drude resistivity and cuts the return off at the total momentum lifetime, but it does
not change the coefficient of the logarithm. With the phonon resistivity known independently, the width scan applies
unchanged, with $\gamma_{\rm tot}(T)$ in place of $\gamma$ in $a_N$.

Several consequences remain to be worked out. A localizing correction cannot be
generated inside the hydrodynamic theory, since it would violate the Wiener bound;
if present it must be inherited from the electronic level. A magnetic field acts on these
returns classically, at the scale of the cyclotron frequency, by mixing the momentum doublet, and not at the
flux-quantum scale of weak localization. The
frequency dependence follows from Eqs.~(\ref{eq:K}) and (\ref{eq:dnut}), including
a $\ln(1/\omega\tau_\xi)$ memory term in the optical conductivity. These, together
with the interplay between the friction return and the long-time tails, the kinetic
wall problem and harmonics $m\ge4$, are left for future work.

\begin{acknowledgments}
The author acknowledges financial support from the Ministerio de Ciencia e
Innovaci\'on through Grant PID2024-161156NB-I00 and the Severo Ochoa Centres of
Excellence program through Grant CEX2024-001445-S. The author used Claude Opus 5.5
(Anthropic) as a research and writing assistant. The author formulated the problem, made the scientific
decisions, checked the results, and takes full responsibility for the content of this article.
\end{acknowledgments}

\section*{Data availability}
The data that support the findings of this article are not publicly available.
The data and the scripts that generate the figures are available from the author upon reasonable request.

\appendix

\section{Kinetic hydrodynamics with uniform relaxation rates}\label{app:kinetic}
\subsection{Boltzmann equation and angular harmonics}
For a two-dimensional isotropic Fermi surface the distribution obeys
\begin{equation}
\begin{gathered}
(\partial_t+\bm v_{\bm p}\cdot\nabla_{\bm x}-e\bm E\cdot\nabla_{\bm p})f=I_{ee}[f],\\
I_{ee}[f_1]=\int_{234}W^{(0)}_{12\to34}\big[f_3f_4(1-f_1)(1-f_2)\\-f_1f_2(1-f_3)(1-f_4)\big],
\end{gathered}
\end{equation}
with $\int_{234}\equiv\int\frac{\dd^2p_2}{(2\pi)^2}\frac{\dd^2p_3}{(2\pi)^2}\frac{\dd^2p_4}{(2\pi)^2}$ and, for
$\hbar=1$, $W^{(0)}_{12\to34}=2\pi|U_{12\to34}|^2\delta(\varepsilon_1+\varepsilon_2-\varepsilon_3-\varepsilon_4)
\delta^{(2)}(\bm p_1+\bm p_2-\bm p_3-\bm p_4)$. Writing $f=f_0-\partial_\varepsilon f_0\,\varphi$ and linearizing,
detailed balance and energy conservation give $I^{\rm lin}_{ee}[\varphi_1]=-C_{ee}[\varphi_1]$ with the positive
operator
\begin{equation}
\begin{gathered}
C_{ee}[\varphi_1]=\int_{234}W_{12\to34}\,(\varphi_1+\varphi_2-\varphi_3-\varphi_4),\\
W_{12\to34}=\beta W^{(0)}_{12\to34}f^0_1f^0_2(1-f^0_3)(1-f^0_4).
\end{gathered}
\label{eq:Cee}
\end{equation}
On the Fermi circle, $\bm v_{\bm p}=v\hat n(\theta)$ with $\hat n=(\cos\theta,\sin\theta)$, we expand
$\varphi=\sum_ma_m(t,\bm x)e^{\ii m\theta}$. With $\partial_\pm=\partial_x\pm\ii\partial_y$ the streaming term
becomes a nearest-neighbour coupling of harmonics,
\begin{equation}
\begin{gathered}
\hat n\cdot\nabla=\tfrac12\big(e^{\ii\theta}\partial_-+e^{-\ii\theta}\partial_+\big),\\
\hat n\cdot\nabla\varphi=\tfrac12\sum_m\big(\partial_-a_{m-1}+\partial_+a_{m+1}\big)e^{\ii m\theta},
\end{gathered}
\end{equation}
and $\partial_+\partial_-=\nabla^2$. Rotational invariance makes the collision operator diagonal,
$C_{ee}[e^{\ii m\theta}]=\gamma^{(0)}_me^{\ii m\theta}$, which gives Eq.~(\ref{eq:hier}) with $\gamma_m$ the
relaxation rate of harmonic $m$. Without quasiparticles the labels $m=0,\pm1,\pm2$ denote the rotation-projected
scalar, vector and spin-two sectors of the hydrodynamic equations; nothing below uses more than the structure of
Eq.~(\ref{eq:hier}).

\subsection{Conservation laws and the two rates}
The rates follow from the positive quadratic form of Eq.~(\ref{eq:Cee}). For $\chi_m=e^{\ii m\theta}$,
\begin{equation}
\begin{aligned}\gamma^{(0)}_mN_m=\frac14\int_{1234}W_{12\to34}\,\big|&\chi_m(\theta_1)+\chi_m(\theta_2)\\&-\chi_m(\theta_3)-\chi_m(\theta_4)\big|^2\ \ge0,\end{aligned}
\end{equation}
so a rate vanishes only if the bracket vanishes wherever the kernel has support. For $m=0$ the bracket is
$1+1-1-1=0$: number is conserved. For $m=\pm1$, $\chi_{\pm1}=(p_x\pm\ii p_y)/p_F$, and the bracket is proportional to
$\bm p_1+\bm p_2-\bm p_3-\bm p_4=0$: momentum is conserved. Momentum conservation does not constrain the quadratic
combination $\chi_2=(p_x+\ii p_y)^2/p_F^2$. For the centre-of-mass process $\theta=(0,\pi,\alpha,\alpha+\pi)$ the
$m=0,1$ brackets vanish while
\begin{equation}
\begin{gathered}
\Delta_2(\alpha)=e^{2\ii\theta_1}+e^{2\ii\theta_2}-e^{2\ii\theta_3}-e^{2\ii\theta_4}=2\big(1-e^{2\ii\alpha}\big),\\
|\Delta_2(\alpha)|^2=16\sin^2\alpha ,
\end{gathered}
\end{equation}
so the stress rate $\Gamma_2\equiv\gamma^{(0)}_{\pm2}$ is positive for any kernel with weight at non-forward angles.
A forward-peaked kernel $K(\alpha)\sim|\alpha|^{-s}$ is softened to $|\alpha|^{2-s}$: forward scattering may dominate
the quasiparticle lifetime without producing a comparable stress-relaxation rate.

Impurities, phonons or umklapp relax momentum at a rate $\gamma\equiv\gamma_{\pm1}\ll\Gamma_2$. The two rates have
different status. $\Gamma_2$ is a coupling of the clean hydrodynamic fixed point. $\gamma$ is not: it multiplies an
operator absent at the fixed point, it parametrizes the explicit breaking of translation invariance, and its value is
set by the sample~\cite{HLS2018}. How disorder can renormalize each of them is the subject of the following appendices.

\subsection{Hydrodynamic reduction}
For $|m|\ge2$ the hierarchy reads $(\partial_t+\gamma_m)a_m=-\frac v2(\partial_-a_{m-1}+\partial_+a_{m+1})$. At
frequencies $\omega\ll\gamma_m$ and wavevectors $q\ell\ll1$, with $\ell=v/\Gamma_2$, the fast harmonics follow the
slow ones,
\begin{equation}
\begin{gathered}
a_m=-\frac{v}{2\gamma_m}\big(\partial_-a_{m-1}+\partial_+a_{m+1}\big)+O\Big(\frac{\omega}{\gamma_m}a_m\Big),\\
a_{\pm m}=O\big((q\ell)^{|m|-1}\big)\,a_{\pm1}\quad(|m|\ge2),
\end{gathered}
\end{equation}
so truncating at $a_{\pm3}=0$ is Navier--Stokes order (the $m=\pm3$ corrections are discussed in
Appendix~\ref{app:window}). The $m=2$ row then gives $a_2=-\frac v2\partial_-a_1/(s+\Gamma_2)$ in Laplace space, and the
$m=1$ row becomes
\begin{equation}
(\partial_t+\gamma)a_1+\frac v2\partial_-a_0-s_1=\nu(s)\nabla^2a_1,\qquad \nu(s)=\frac{v^2/4}{s+\Gamma_2}.
\end{equation}
With $u_x=\frac v2(a_1+a_{-1})$ and $u_y=\frac{\ii v}{2}(a_1-a_{-1})$ the slow sector is the linearized
Navier--Stokes system with friction,
\begin{equation}
\partial_ta_0+\nabla\cdot\bm u=0,\qquad
(\partial_t+\gamma)\bm u+\frac{v^2}{2}\nabla a_0-\frac{ev^2}{2}\bm E=\nu\nabla^2\bm u ,
\end{equation}
with $\nu=v^2/4\Gamma_2$. A correction to $\Gamma_2$ is inherited by the viscosity,
$\delta\nu/\nu=-\delta\Gamma_2/\Gamma_2$ to first order.

\subsection{Longitudinal and transverse channels}
In Fourier space the hierarchy matrix has $M_{mm}=s+\gamma_m$ and $M_{m,m\mp1}=(\ii vq/2)e^{\mp\ii\varphi}$, and
the rows $m=0,\pm1$ are continuity and momentum balance, with $n=a_0$, $j_x=v(a_1+a_{-1})/2$ and
$j_y=\ii v(a_1-a_{-1})/2$. Rotating to the frame of $\bm q$, $b_m=e^{\ii m\varphi}a_m$, and forming the even and odd
combinations $e_k=(b_k+b_{-k})/\sqrt2$, $o_k=(b_k-b_{-k})/\sqrt2$ splits it into a longitudinal and a transverse block,
\begin{equation}
\begin{gathered}
M_L=\begin{pmatrix}s&\ii vq/\sqrt2&0\\ \ii vq/\sqrt2&s+\gamma&\ii vq/2\\0&\ii vq/2&s+\Gamma_2\end{pmatrix},\\
M_T=\begin{pmatrix}s+\gamma&\ii vq/2\\ \ii vq/2&s+\Gamma_2\end{pmatrix}.
\end{gathered}
\end{equation}
Each harmonic amplitude has weight $\frac12$ in each channel, so the stress propagator is
$G_{2,2}=\frac12(G_L+G_T)$ and $G_{2,-2}=\frac12e^{-4\ii\varphi}(G_L-G_T)$. The momentum propagators are
\begin{equation}
\begin{gathered}
G^T_{11}=\frac{1}{s+\gamma+\nu(s)k^2},\\
G^L_{11}=\frac{s}{s(s+\gamma)+(c^2+s\nu(s))k^2},\qquad c^2=\frac{v^2}{2},
\end{gathered}
\end{equation}
and the stress line obeys the exact resolvent identity
\begin{equation}
G_{22}(s,k)=\frac{1}{s+\Gamma_2}-\frac{v^2k^2}{4}\,\frac{G_{11}(s,k)}{(s+\Gamma_2)^2}.
\label{eq:resolvent}
\end{equation}
At DC the longitudinal momentum propagator vanishes identically. Physically, continuity forces
$\bm q\cdot\bm u=0$ in a steady state, and a static longitudinal force builds up a pressure gradient rather than a
flow. All DC transport is therefore carried by the transverse (shear) channel, $[M_T^{-1}(0)]_{11}=1/(\gamma+\nu
k^2)$, in which the viscosity appears only after the stress harmonic is eliminated. The clean fixed point has sound
speed $c=v/\sqrt2$ and sound damping $\nu q^2/2$. Both stress branches sit at
$s=-\Gamma_2+v^2k^2/[4(\Gamma_2-\gamma)]+O(k^4)$, so $\Sigma_{22}(s)$ is non-analytic at $s=-\Gamma_2$ and the
physical viscosity is the DC value $\nu_{\rm eff}(s\to0)$.

\section{Disorder on top of the momentum-relaxation rate}\label{app:admissible}
\subsection{Which local disorder is allowed}
We ask which static, local perturbations $V(\bm x)$ of the hierarchy, acting on $a_m$ with $|m|\le2$ and without
derivatives, describe disorder \emph{on top of} $\gamma$. Five physical requirements restrict them: (R) the fields are
real, $JVJ=V^*$, with $J$ the exchange $m\leftrightarrow-m$; (N) number is conserved and the uniform equilibrium is stationary, so the row and column $m=0$
vanish; (T) Onsager reciprocity, $V^\dagger=\mathcal{T}V\mathcal{T}$, with $\mathcal{T}={\rm diag}[(-1)^m]$ the momentum reversal $\bm p\to-\bm p$; (S) the dissipative part stays positive,
$R+H(\bm x)\succeq0$ on $m\neq0$, with $R={\rm diag}(\Gamma_2,\gamma,0,\gamma,\Gamma_2)$ and $H=(V+V^\dagger)/2$; and
(Iso) the ensemble is statistically isotropic, so an entry connecting harmonics $k$ steps apart requires a random field
of angular momentum $k$. These conditions cut the 25 real parameters of a general $5\times5$ vertex to 16 (R, N), 10
(R, N, T), and finally 2 (all five): $V={\rm diag}(\delta\Gamma_2,\delta\gamma,0,\delta\gamma,\delta\Gamma_2)$.
Microreversible elastic impurity scattering spans all ten directions allowed by (R, N, T), and its isotropic part lands
exactly on $\{\delta\gamma,\delta\Gamma_2\}$, which is an independent check on the count.

The momentum Ward identity then sorts the perturbations into classes according to the integrated momentum balance at
$q=0$, $\frac{\dd}{\dd t}\int\bm j=-\int(\text{rows }\pm1\text{ of }R+V)\,a$:
\begin{itemize}
\item W0, rows $\pm1$ vanish: $\delta\Gamma_2$, an anisotropic stress relaxation, and an odd viscosity. Momentum then
relaxes at exactly $\gamma$ in every realization.
\item W1-dissipative: $\delta\gamma$ and a traceless friction tensor. Positivity gives $\delta\gamma\ge-\gamma$ and
bounds the tensor by $\gamma+\delta\gamma$, so these perturbations are \emph{slaved} to $\gamma$ and vanish with it.
\item W1-reversible: momentum--stress couplings and a random field acting on momentum. They are not bounded by
positivity, and at second order they relax momentum even at $\gamma=0$. They are independent symmetry-breaking
parameters and are excluded by our premise.
\end{itemize}
We use the minimal isotropic ensemble $\delta\gamma=\gamma w_1$, $\delta\Gamma_2=\Gamma_2w_2$, i.e. a random friction
and a random stress-relaxation rate, equivalently a random viscosity $\nu(\bm x)=\nu/(1+w_2)$, with
$\langle w_a(\bm x)w_a(\bm x')\rangle=\sigma_a^2e^{-|\bm x-\bm x'|^2/2\xi^2}$ and
$C_a(k)=2\pi\sigma_a^2\xi^2e^{-k^2\xi^2/2}$.

Two natural alternatives are excluded, for different reasons. An unbounded friction ensemble, for
instance a Gaussian friction $\zeta(\bm x)$ of mean $\bar\zeta$ and variance $\Delta_0$, violates positivity. Already in
zero dimensions $\langle e^{-\zeta t}\rangle=e^{-\bar\zeta t+\Delta_0t^2/2}$, which grows for $t>2\bar\zeta/\Delta_0$:
the antidamping realizations dominate the average, and the averaged self-energy acquires the wrong sign. The physical
friction is $\gamma(1+w_1)$ with $1+w_1\ge0$, whose mean is the breaking parameter itself; the Gaussian $w_1$ used at
Born order stands for a positive ensemble with the same covariance. A reversible perturbation, for instance a static
random rotation $\partial_t\bm u=\xi(\bm x)\,\hat z\times\bm u$ (the graded vertex $\ii\xi\,{\rm sgn}(m)$, whose odd
grading is forced by reality), does no work in any realization, yet relaxes momentum at second order even at
$\gamma=0$: it belongs to the W1-reversible class.

A density landscape, $n=\bar n(1+w)$, enters continuity as $\nabla\cdot(\delta n\,\bm u)$ and is an independent
parameter $\sigma_n$. At second order its cross term with $w_1$ cancels for any cross-correlation, so it adds to the
friction and stress results without changing them; it is treated in Appendix~\ref{app:aks}.

\subsection{Two exact consequences of the Ward identity}
\emph{Confinement.} Writing $T=V(1+G_0V)^{-1}$, the self-energy is
$\Sigma=-\langle T\rangle(1-G_0\langle T\rangle)^{-1}$. Every term begins and ends on a disorder vertex, so
$\Sigma=P\Sigma P$ to all orders and for any statistics, with $P$ the projector on $m=\pm1,\pm2$. At $q=0$ stress
disorder therefore leaves the momentum line exactly unchanged, $\langle G\rangle_{11}=1/(s+\gamma)$, and friction
disorder leaves the stress line exactly unchanged, $\langle G\rangle_{22}=1/(s+\Gamma_2)$. Each kind of disorder is
confined to its own rate.

\emph{Wiener bounds.} On a lattice of cells with local rates $\gamma_i$, the integrated momentum
balance $\sum_i\gamma_iX_i=NF$ and the energy balance with positivity, $\sum_i\gamma_iX_i^2\le F\sum_iX_i$, give
\begin{equation}
\langle\gamma(\bm x)^{-1}\rangle^{-1}\le\gamma_R\le\langle\gamma(\bm x)\rangle=\gamma .
\end{equation}
The upper bound fixes the sign: no static disorder of this kind can increase momentum relaxation. In weak
localization the sign is set by the symmetry class; here it is fixed by the Ward identity together with positivity.

\subsection{Disorder average and power counting}
Static Gaussian disorder can be averaged exactly in the Martin--Siggia--Rose formulation. With weight $e^{-S}$,
$S=\int\tilde a(\partial_t+L_0+R+V)a$, the average generates
\begin{equation}
\begin{aligned}S_{\rm dis}=-\frac12\sum_{AB}R_AR_B\!\int\!\dd t\,\dd t'\,\dd^2x\,\dd^2x'\,&C_{AB}(\bm x-\bm x')\\
\times[\tilde aO_Aa]_{x,t}&[\tilde aO_Ba]_{x',t'},\end{aligned}
\end{equation}
local in space and bilocal in time because the disorder is static. Two consequences matter. First, static disorder
transfers momentum but no energy, so the self-energy is $\Sigma(\ii\omega_n)=\langle VG(\ii\omega_n)V\rangle$ with no
internal frequency sum; inserting one, or using the classical Keldysh component of the fluctuating velocity in place of
the retarded response, produces a spurious factor of $T$. Second, at the viscous fixed point ($z=2$, $d=2$) the friction
vertex is marginal relative to $\gamma$, with dimensionless coupling $2\pi\sigma_1^2(\xi/D_\nu)^2$, while the stress
vertex is irrelevant, with scaling dimension $-2$ and coupling $2\pi\sigma_2^2(\xi/\ell)^2$. Rescaling the action,
counting units, and the Harris criterion give the same dimensions, and reproduce the known random-mass ($4-d$) and
random-diffusion ($-d$) results.

The origin of the logarithm is visible in two one-loop integrals. The slaved friction vertex is a constant, $\gamma$,
and the transverse momentum line has mass $\gamma$:
\begin{equation}
\begin{gathered}
\int_0^\infty\frac{k\,\dd k}{2\pi}\,\frac{\gamma^2C(k)}{\gamma+\nu k^2}=\gamma\sigma^2f(a_0),\qquad a_0=\frac{\gamma\xi^2}{2\nu},\\
\gamma\sigma^2f(a_0)=\gamma\sigma^2a_0\Big[\ln\frac1{a_0}-\gamma_E\Big]+O(a_0^2\ln a_0),
\end{gathered}
\end{equation}
a marginal logarithm. A vertex constrained by an exact conservation law is instead a total divergence, $V\propto uk$ on
the conserved density with propagator $1/(s+Dk^2)$, and
\begin{equation}
\lim_{s\to0}\int_0^\infty\frac{k\,\dd k}{2\pi}\,\frac{(uk)^2C(k)}{s+Dk^2}=\frac{u^2\sigma^2}{D}
\label{eq:Sfinite}
\end{equation}
is finite. In both integrals the correlator cuts the ultraviolet at $k\sim1/\xi$; the infrared is cut by $\gamma$ in the
first and needs no cut in the second. A two-dimensional return logarithm therefore requires both a soft mode and a
gradient-free coupling to it.

\section{Born order}\label{app:born}
\subsection{Friction: the return kernel}
At second order in $w_1$ the transverse and longitudinal momentum self-energies follow from the radial integral
$\int_0^\infty\dd u\,e^{-u\xi^2/2}/(p+nu)=e^aE_1(a)/n$, $a=p\xi^2/2n$:
\begin{equation}
\begin{gathered}
\Sigma^T_{11}(s)=\frac{\gamma^2\sigma_1^2\xi^2}{4\nu(s)}\,e^{A}E_1(A),\qquad A=\frac{(s+\gamma)\xi^2}{2\nu(s)},\\
\Sigma^L_{11}(s)=\frac{\gamma^2\sigma_1^2\xi^2}{4}\,\frac{s}{c^2+s\nu(s)}\,e^BE_1(B).
\end{gathered}
\end{equation}
The longitudinal piece vanishes at DC, as it must. For $s\ll\Gamma_2$ the transverse self-energy is the Laplace
transform of a memory kernel with a transparent structure,
\begin{equation}
\begin{gathered}
K(t)=\frac{\gamma^2\sigma_1^2\xi^2\,e^{-\gamma t}}{2(\xi^2+2\nu t)}=\frac{\gamma^2}{2}C_1(0)\,P_\nu(t)\,e^{-\gamma t},\\
P_\nu(t)=\int\!\frac{\dd^2k}{(2\pi)^2}e^{-k^2\xi^2/2}e^{-\nu k^2t}=\frac{1}{4\pi\nu(t+\tau_\xi)}:
\end{gathered}
\end{equation}
the square of the slaved vertex, the disorder strength at zero wavevector, and the two-dimensional return probability
of transverse momentum diffusing with $\nu$, cut off at short times by $\tau_\xi=\xi^2/2\nu$ and at long times by
$1/\gamma$. Its time integral gives the renormalized rate
\begin{equation}
\begin{gathered}
\frac{\gamma_R}{\gamma}=1-\frac{\sigma_1^2}{2}f(a_0),\qquad a_0=\gamma\tau_\xi,\\
f(a)=a\,e^aE_1(a)=\int_0^\infty\dd y\,e^{-y}\frac{a}{a+y}\in(0,1),
\end{gathered}
\end{equation}
and in the frequency domain
$\Sigma^T_{11}(s)\simeq(\gamma^2C_1(0)/8\pi\nu)[\ln(1/(s+\gamma)\tau_\xi)-\gamma_E]$. For an arbitrary isotropic
correlator the Born results take the exact form $\delta\gamma/\gamma=-I_1$ and $\delta\nu/\nu=\sigma_2^2/2+I_2$, with
$I_a=(\gamma/2\nu)\int_0^\infty\frac{k\,\dd k}{2\pi}C_{aa}(k)/(k^2+\gamma/\nu)$. In real space, with
$\int_0^\infty k\,\dd k\,J_0(kr)/(k^2+D_\nu^{-2})=K_0(r/D_\nu)$,
\begin{equation}
\begin{gathered}
I_a=\frac{1}{4\pi D_\nu^2}\int\dd^2r\,\langle w_a(\bm0)w_a(\bm r)\rangle\,K_0(r/D_\nu)\\
=\frac{C_a(0)}{4\pi D_\nu^2}\Big[\ln\frac{2D_\nu}{\rho_c}-\gamma_E\Big]+O\Big(\frac{\rho_c^2}{D_\nu^2}\ln\frac{D_\nu}{\rho_c}\Big),
\end{gathered}
\end{equation}
for correlations of range much shorter than $D_\nu$, where $K_0(r/D_\nu)=\ln(2D_\nu/r)-\gamma_E+O[(r^2/D_\nu^2)\ln(D_\nu/r)]$ has been
used. Here $C_a(0)=\int\dd^2r\,\langle w_a(\bm0)w_a(\bm r)\rangle$ and $\ln\rho_c$ is the mean of $\ln r$ weighted by
$\langle w_a(\bm0)w_a(\bm r)\rangle\,\dd^2r$, the mean log-distance of the correlator that also enters the narrow-channel
logarithm of Appendix~\ref{app:channel}. The coefficient of the logarithm therefore depends only on $C_a(0)$, and the
shape of the correlator enters only through $\rho_c$: $\rho_c=\sqrt2\,e^{-\gamma_E/2}\xi$ for the Gaussian, which
reproduces Eq.~(\ref{eq:gammaR}), and $\rho_c=e^{1-\gamma_E}\ell$ for an exponential correlator $e^{-r/\ell}$.

Three results obtained by other routes agree with these: a two-scale (Brinkman) homogenization reproduces the DC
values; in the local limit $a_0\to\infty$ the correction reduces to the two-dimensional Darcy effective-medium value
$\gamma_{\rm eff}=\gamma(1-\langle w^2\rangle\langle P^T_{xx}\rangle)$ with $\langle P^T_{xx}\rangle=\frac12$; and the
results lie inside the Wiener and Hill bounds~\cite{Torquato2002}.

\subsection{Stress: the viscosity correction}
At $q=0$ the stress self-energy is
\begin{equation}
\begin{gathered}
\Sigma_{22}(s)=\Gamma_2^2\sigma_2^2\xi^2\Big\{\frac{1}{\xi^2(s+\Gamma_2)}-\frac{v^2\big[K_1^T+sK_1^L\big]}{8(s+\Gamma_2)^2}\Big\},\\
K_1=\frac{1/\xi^2-PK_0}{N},\qquad K_0=\frac{e^{P\xi^2/2N}E_1(P\xi^2/2N)}{2N},
\end{gathered}
\end{equation}
with $(P,N)=(s+\gamma,\nu(s))$ in the transverse and $(s(s+\gamma),c^2+s\nu(s))$ in the longitudinal channel. The DC
value has a simple reading. The internal stress line is $G_{22}(0,k)=(1+h(k))/2\Gamma_2$ with
$h=\gamma/(\gamma+\nu k^2)$: its longitudinal half is local, $1/\Gamma_2$, while its transverse half converts into the
shear mode, $h/\Gamma_2$. Hence $\Sigma_{22}(0)=(\Gamma_2\sigma_2^2/2)(1+f)$ and
\begin{equation}
\frac{\delta\nu}{\nu}=-\frac{\delta\Gamma_2}{\Gamma_2}=\frac{\sigma_2^2}{2}\big[1+f(a_0)\big].
\end{equation}
The viscosity is raised. The constant is a local (Voigt) rescaling; the temperature dependence enters through
$f(a_0)$, generated by the same return of transverse momentum as the friction correction.

Eliminating the stress row gives $\nu_R(s)=(v^2/4)/(s+\Gamma_2-\Sigma_{22}(s))$, so
$\delta\nu(s)\to(\nu/\Gamma_2)\Sigma_{22}$ for $s\ll\Gamma_2$. Using
$\int_0^\infty p^3e^{-p^2(\xi^2/2+\nu t)}\dd p=1/[2\nu^2(t+\tau_\xi)^2]$ for the transverse part, the viscosity memory
kernel is
\begin{equation}
\begin{gathered}
\delta\nu(t)=\nu\sigma_2^2\Big[\Gamma_2 e^{-\Gamma_2 t}-\frac{\tau_\xi}{2}\frac{e^{-\gamma t}}{(t+\tau_\xi)^2}\Big],\\
\frac{\delta\nu(s)}{\nu}=\frac{\sigma_2^2}{2}\Big[1+f\big((s+\gamma)\tau_\xi\big)\Big]\quad(s\ll\Gamma_2),
\end{gathered}
\end{equation}
whose Laplace transform, via $\int_0^\infty e^{-bt}(t+\tau)^{-2}\dd t=1/\tau-be^{b\tau}E_1(b\tau)$, reproduces the
frequency-domain result. Three independent facts are consistent with it: at $\gamma=0$ the tail integral is exactly
$\frac12$, the coefficient required at this order by the exact two-dimensional Stokes duality
(Appendix~\ref{app:beyond}); for $a_0\to\infty$ the tail is bounded by $1/2a_0$, leaving the Voigt mean
$\langle\omega\rangle=\sigma_2^2$; and the $t^{-2}$ tail matches the scaling dimension $-2$ of the stress vertex. A
longitudinal (sound) transient of duration $\sim\xi/c$ and zero time integral is present at short times; it does not
affect the DC results.

To second order in the external wavevector,
$\Sigma^T_{11}(0,q)=(\gamma\sigma_1^2/2)f+q^2(\gamma^2\sigma_1^2\xi^4/2\nu)[3/8-(1/4+3a/8)E]$ and
$\Sigma_{22}(0,q)=(\Gamma_2\sigma_2^2/2)(1+f)+q^2(\Gamma_2\sigma_2^2\xi^2a/4)[1-(1+a)E]$, with $E=e^aE_1(a)$. The
friction contribution to the viscosity, $\delta\nu_{\rm fric}/\nu=(\sigma_1^2a_0^2/2)[\ln(1/a_0)-\gamma_E-3/2]$, is one
power of $a_0$ below the terms kept. A trace sum rule in the Darcy limit fixes the transverse and longitudinal weights
$3/8$ and $1/8$.

\subsection{Mode-space diagrams and the transport equation}
The diagrams of Fig.~\ref{fig:diagrams} use bare propagators $1/s$ ($m=0$), $1/(s+\gamma)$ ($m=\pm1$) and
$1/(s+\Gamma_2)$ ($m=\pm2$), streaming vertices $(\ii vq/2)e^{\mp\ii\varphi}$, and disorder vertices $-\gamma w_1$
on $m=\pm1$ and $-\Gamma_2w_2$ on $m=\pm2$. The self-energy is ${\rm diag}(0,\Sigma_{11},\Sigma_{22})$: no vertex touches
$m=0$, and the two disorders are uncorrelated. Eliminating the stress and density rows,
\begin{equation}
\begin{gathered}
(\bar G_T)_{11}^{-1}=s+\gamma-\Sigma_{11}+\frac{v^2k^2/4}{s+\Gamma_2-\Sigma_{22}},\\
(\bar G_L)_{11}^{-1}=(\bar G_T)_{11}^{-1}+\frac{v^2k^2/2}{s},
\end{gathered}
\end{equation}
so only the transverse channel survives at DC. In a single realization the stress row with
$\Gamma_2(\bm x)=\Gamma_2(1+w_2)$ produces exactly $\nu(\bm x)pk$ with $\nu(\bm x)=\nu/(1+w_2)\equiv\nu(1+\omega)$: the
random-viscosity Brinkman problem. Projected on the transverse directions of the two momenta, its first order in
$w_2$ is $\nu\hat\omega\,pk\cos2\theta$, the vertex $\blacksquare$ of Fig.~\ref{fig:diagrams}; the same vertex follows independently
from $-\nabla\cdot(\nu\omega\bm S)$. Its second order at one point is the local (Voigt) insertion
$\langle\omega\rangle\nu p^2=\sigma_2^2\nu p^2$.

Acting with the bare inverse propagator on the Dyson equation and eliminating $m=\pm2$ gives a generalized
Navier--Stokes equation with two memory terms,
\begin{equation}
\begin{aligned}\partial_t\bm u+\nabla P={}&-\gamma\bm u+\int_{-\infty}^t\!\dd t'\,K(t-t')\,\bm u(t')\\
&+\nabla\!\cdot\!\int_{-\infty}^t\!\dd t'\,\nu_R(t-t')\,\bm S[\bm u](t')+\bm F,\end{aligned}
\end{equation}
with $\nabla\cdot\bm u=0$ at DC and
$\nu_R(t)=\nu(1+\sigma_2^2)\Gamma_2e^{-\Gamma_2t}-\nu\sigma_2^2(\tau_\xi/2)e^{-\gamma t}/(t+\tau_\xi)^2$.

\subsection{Different status, different renormalization}
The two Born results make the different status of $\gamma$ and $\Gamma_2$ concrete. The friction self-energy at DC is
\begin{equation}
\begin{gathered}
\Sigma^T_{11}(0)=\frac{\gamma^2}{2}\int\frac{\dd^2k}{(2\pi)^2}\,\frac{C_1(k)}{\gamma+\nu k^2}=\frac{\gamma\sigma_1^2}{2}f(a_0),\\
\frac{\delta\gamma}{\gamma}\simeq-\frac{\sigma_1^2\xi^2}{4\nu}\,\gamma\Big[\ln\frac{2\nu}{\gamma\xi^2}-\gamma_E\Big]\ \xrightarrow{\ \gamma\to0\ }\ 0,
\end{gathered}
\end{equation}
where $\frac12$ is the transverse weight and $\gamma^2$ comes from the two slaved vertices. The breaking parameter is
thus renormalized only multiplicatively and at higher order in itself, with a logarithm of its own size, because both
the vertex and the mass of the mode in the loop are proportional to $\gamma$. The same holds through fourth order in
the disorder: the coefficients $c_4$ and $c_{12}$ of Table~\ref{tab:c4} vanish as $a_0\to0$. The stress-relaxation rate,
by contrast, is shifted already at leading order, $\Sigma_{22}(0)=(\Gamma_2\sigma_2^2/2)[1+f(a_0)]$: its first term is
independent of $\gamma$ and is absorbed into the measured viscosity, while its second term, generated by the return of
transverse momentum, vanishes with $\gamma$. The temperature dependence of the stress correction is therefore
induced by the breaking. Table~\ref{tab:ward} summarizes what each identity allows.

\begin{table*}[t]
\caption{\label{tab:ward}Ward identities, allowed disorder couplings, and the corrections they allow.}
\begin{ruledtabular}\small
\begin{tabular}{llllll}
identity & status & soft mode & disorder coupling & parameter, form & sign\\ \hline
number & exact & density & divergence & none at $q=0$; finite & --\\
momentum (W1) & broken by $\gamma$ & transverse mom. & $\gamma w_1$, no gradient & $\gamma$: $-\frac{\sigma_1^2}{2}a_0\ln\frac1{a_0}$ & $\gamma_R\le\gamma$\\
momentum (W0) & kept at $q=0$ & -- & $\Gamma_2w_2$, two gradients & $\nu$: $\frac{\sigma_2^2}{2}[1+f(a_0)]$ & $\delta\nu>0$\\
time reversal & until $\tau_\varphi$ & Cooperon & no gradient & $\sigma$: $\ln(\tau_\varphi/\tau)$ & sym.\ class\\
\end{tabular}
\end{ruledtabular}
\end{table*}

\subsection{Relation to the Hershfield--Ambegaokar equation}\label{app:HA}
Hershfield and Ambegaokar (HA) wrote weak localization as a memory term in the Boltzmann
equation~\cite{HA1986},
\begin{equation}
\begin{gathered}
\begin{aligned}\relax[\partial_T+v_F\hat p\cdot\nabla]f={}&-\Gamma[f-\bar f]+\int_{-\infty}^T\!\!\dd T'\,\alpha(T-T')\\&\times[f(-\bm p,T')-\bar f(T')],\end{aligned}\\
\alpha(t)=2n_i|V|^2\!\int'\!\frac{\dd^dq}{(2\pi)^d}e^{-(Dq^2+\tau_\varphi^{-1})t},
\end{gathered}
\end{equation}
with $\Gamma=2\pi n_i|V|^2N(0)$ their elastic rate (not the stress rate), $N(0)$ per spin, $|\bm q|<1/\ell$, and
$d=3$. The dimension enters only through $D=v^2/d\Gamma$ and the $q$ measure: the local bubble
$\int\dd\varepsilon/[(\omega-\varepsilon)^2+\Gamma^2/4]=2\pi/\Gamma$ and the Cooperon $u\Gamma/(Dq^2-\ii\Omega)$ are
dimension independent. In $d=2$, $\int\alpha/\Gamma=\ln(\tau_\varphi/\tau)/(\pi k_F\ell)$ and $\sigma\propto1/(\Gamma+\alpha)$
reproduce $\delta\sigma=-(e^2/2\pi^2\hbar)\ln(\tau_\varphi/\tau)$~\cite{GLK1979}; a single-rung normalization would miss
it by a factor of 2. Projected on $m=1$, HA's term gives $\partial_ta_1=-(\Gamma+\alpha)a_1$ (localizing), whereas ours
gives $\partial_t\bm u=-(\gamma-\int K)\bm u$ (anti-localizing). The normalizations also differ: HA's memory is twice the
rung (maximally crossed, unit residue), ours half the rung (a Born rainbow with the transverse projector), so
importing HA's rule here would overcount by a factor of 4.

\section{The Cooperon in the hydrodynamic window}\label{app:cooperon}
In a two-dimensional Fermi liquid at intermediate
temperatures, $T\tau_{\rm mr}\gg1$, electron--electron collisions dephase at
$1/\tau_\varphi\sim(T^2/E_F)\ln(E_F/T)$~\cite{NZA2002}, while the stress-relaxation rate is $\Gamma_2\sim T^2/E_F$ up
to a number~\cite{LGL2019}. Hence $\tau_\varphi\lesssim1/\Gamma_2$ and
\begin{equation}
\frac{\tau_\varphi}{\tau_{\rm mr}}\lesssim\frac{\gamma}{\Gamma_2}\ll1 :
\end{equation}
electrons are dephased many times before impurity scattering can close a loop. The diffusive Cooperon logarithm
requires $\tau_\varphi\gg\tau_{\rm mr}$ and is absent; at most a ballistic, log-free remainder of order
$1/k_F\ell_{\rm mr}$ survives, and the classical friction return is the only return correction with a logarithm inside
the window. This is a parametric estimate, not a calculation. The order-one prefactors are not derived (realistic
$\Gamma_2/\gamma\sim10$--$100$ gives $\tau_\varphi/\tau_{\rm mr}\sim0.01$--$0.1$), the Planckian case is a plausibility
argument, and narrow channels and interaction corrections of the Zala--Narozhny--Aleiner type are not addressed.

In kinetic form, coherent backscattering is a gain term at the reversed momentum with a compensating uniform loss,
$\delta I_{\rm coh}[f]=\alpha[f(-\bm p)-\bar f]$. Since $f(\theta+\pi)$ multiplies harmonic $m$ by $(-1)^m$,
\begin{equation}
\Gamma^{\rm tot}_m=\gamma_m-(-1)^m\alpha_m\quad(m\neq0),\qquad \Gamma^{\rm tot}_0=0 :
\end{equation}
the momentum rate is enhanced (weak localization), and a stress rate would be reduced. This is only a kinematic
projection; the amplitudes $\alpha_m$ in a fluid are not computed here.

Were both present, they would add. The Cooperon is off-diagonal in time reversal and acts on the electronic
propagator; the classical return is diagonal and acts on the transverse momentum propagator. They carry opposite
signs in the momentum sector and different temperature and field dependences: dephasing and the flux quantum for the
first, $\nu(T)$, $\gamma$ and the cyclotron frequency for the second.

\section{Beyond Born order}\label{app:beyond}
At DC and with $a_{\pm3}=0$ the kinetic problem reduces exactly to the random Brinkman problem,
Eq.~(\ref{eq:brinkman}). We expand it to fourth order in the disorder amplitudes in order to answer three
questions: whether the logarithm exponentiates or needs resummation, how large the disorder can be before the Born
result fails, and whether new logarithmic structures appear. Below we use units $\xi=1$, scaled squared radii
$u_x=|\bm x|^2/2$ in momentum space, and $h_x=a_0/(a_0+u_x)$.

\subsection{Friction to fourth order}
Removing the reducible pairing, $\Sigma_4=\langle T\rangle_4+\langle T\rangle_2G\langle T\rangle_2$, gives
\begin{equation}
\begin{gathered}
\frac{\gamma_{\rm eff}}{\gamma}=1-\frac{\sigma^2}{2}f+K_3-(R+X),\\ R=\tfrac12\sigma^4E[h_p^2h_q(\hat p\cdot\hat q)^2],\\
X=\tfrac12\sigma^4E[h_ph_{p'}h_q(\hat p'\cdot\hat q)(\hat q\cdot\hat p)(\hat p\cdot\hat p')],
\end{gathered}
\end{equation}
with $\bm q=\bm p+\bm p'$, $E$ the Gaussian average over the scaled momenta, and hub terms $K_3=0$ for Gaussian
disorder. In the local (Darcy) limit the pointwise identity
$(\bm p\cdot\bm q)^2|\bm p'|^2-(\bm p'\cdot\bm q)(\bm q\cdot\bm p)(\bm p\cdot\bm p')=(\bm p\cdot\bm q)(\bm p\times\bm p')^2$
gives $R'-X'=\frac14$ for any isotropic law. For a lognormal medium this reproduces
$1-\sigma^2/2+3\sigma^4/8=(1+\sigma^2)^{-1/2}$, the exact result of Dykhne~\cite{Dykhne1971,Matheron1967}; the
Keller--Dykhne reciprocity~\cite{Keller1964} requires the same identity through different combinatorics. For Gaussian
disorder the Darcy coefficient is $c_4^{\rm Darcy}=-(7/8-\frac12\ln2)=-0.528426$. Table~\ref{tab:c4} gives $c_4(a_0)$
across the crossover. For small $a_0$,
\begin{equation}
c_4=-\tfrac14a_0^2\big[\ln(1/a_0)-\gamma_E\big]-X_0a_0^2,\qquad X_0\approx0.47,
\end{equation}
a single logarithm: the series in $\sigma^2$ does not generate powers of $\ln(1/a_0)$ that would require
resummation. Rigorous bounds, $0\le R\le(a_0^2/2)[3+\ln(1+1/a_0)]$ and $|X|\le3.76\,a_0^2$, follow from
$E_\theta[h(\bm q)]=a_0/S\le\min(1,a_0/|x-y|)$ with $S^2=(x-y)^2+a_0^2+2a_0(x+y)$ and $E_1(a)<\ln(1+1/a)$.

\begin{table*}[t]
\caption{\label{tab:c4}Fourth-order coefficients across the crossover: friction $c_4$, mixed friction--stress $c_{12}$, and
stress $c_4^\nu$, defined by $\gamma_{\rm eff}/\gamma=1-\frac{\sigma_1^2}{2}f+\sigma_1^4c_4+\sigma_1^2\sigma_2^2c_{12}$ and
$\delta\nu/\nu=\frac{\sigma_2^2}{2}(1+f)+\sigma_2^4c_4^\nu$ (quadrature of the exact angular reductions).}
\begin{ruledtabular}
\begin{tabular}{lccccccccc}
$a_0$ & $10^{-3}$ & $10^{-2}$ & $0.1$ & $0.3$ & $1$ & $3$ & $10$ & $100$ & $\infty$\\ \hline
$c_4$ & $-2.017\times10^{-6}$ & $-1.336\times10^{-4}$ & $-0.005803$ & $-0.025907$ & $-0.095212$ & $-0.21790$ & $-0.36857$ & $-0.50562$ & $-0.528426$\\
$c_{12}$ & $1.3323\times10^{-3}$ & $7.9551\times10^{-3}$ & $0.034194$ & $0.054297$ & $0.066878$ & $0.056904$ & $0.031823$ & $0.0047225$ & $0$\\
$c_4^\nu$ & -- & $0.70736$ & $0.91802$ & $1.18415$ & $1.65913$ & $2.17622$ & $2.62805$ & $2.95183$ & $3$\\
\end{tabular}
\end{ruledtabular}
\end{table*}

\subsection{Friction dressed by stress disorder}
The stress vertex vanishes on a zero-momentum leg, so to $O(\sigma_1^2)$ and all orders in $\sigma_2$ the friction
correction is $\gamma_{\rm eff}/\gamma=1-(\gamma/2)\int C_1\bar g_T$, with $\bar g_T$ the shear propagator averaged over
the stress disorder. Writing its viscosity as $\delta\nu(p)/\nu=\sigma_2^2[1-J(u)]$, with
$J(u)=\int_0^\infty\dd v\,e^{-(u+v)}\frac{v}{a_0+v}\frac12[I_0+I_4](2\sqrt{uv})$, the mixed coefficient is
\begin{equation}
\begin{gathered}
c_{12}(a_0)=\tfrac12E_u[h(1-h)(1-J(u))],\\
c_{12}=\tfrac14(1+f)\big(f-a_0(1-f)\big)+c_Ma_0+O(a_0^2\ln a_0),\\ c_M=\tfrac{11}{8}-2\ln2 .
\end{gathered}
\end{equation}
The first term is what a self-consistent Born treatment would give; $c_Ma_0$ is the correction to it. The small-$a_0$
form was obtained by a termwise sum and, independently, by Mellin continuation, and it respects the Voigt--Reuss
bracket $0\le c_{12}\le\frac12E[h(1-h)]$ and the Darcy limit $c_{12}\to0$.

\subsection{Stress to fourth order and the Stokes duality}
With the cumulants of $\omega=1/(1+w_2)-1$,
\begin{equation}
\begin{gathered}
\frac{\delta\nu}{\nu}=\frac{\sigma_2^2}{2}(1+f)+\sigma_2^4c_4^\nu,\\
c_4^\nu=f+f(a_0/2)+\tfrac12a_0(1-f)+\tfrac12f^2+2\tilde R-R_2-X_2,
\end{gathered}
\end{equation}
where $\tilde R$, $R_2$ and $X_2$ are Gaussian averages of products of $(1-h)$ factors and doubled-angle cosines over
$\bm p$, $\bm p'$ and $\bm s=\bm p+\bm p'$. The doubled-angle identity $F(\bm p,\bm p')+F(\bm p',\bm p)=\sin^22\theta$
plays the role of the friction identity and gives $c_4^\nu(0)=\frac{33}{8}-5\ln2=0.6593$; the local limit is the Voigt
value $c_4^\nu(\infty)=3$. An exact two-dimensional Stokes duality checks both. A $90^\circ$ rotation in
traceless-strain space maps compatible strain rates onto equilibrated deviatoric stresses, so
$\nu_{\rm eff}[\nu]\,\nu_{\rm eff}[\nu_0^2/\nu]=\nu_0^2$. At $O(\sigma^4)$ this requires the same identity, and for a
lognormal viscosity it gives the geometric mean.

\subsection{Logarithms at fourth order}
Two new logarithmic structures appear at fourth order. First, the stress coefficient carries a subleading logarithm,
$c_4^\nu(a_0)-c_4^\nu(0)=a_0[\alpha\ln(1/a_0)+\beta]+o(a_0)$ with $\alpha=5/4$. Terms with two or more factors of
$h$ are $O(a_0)$ without a logarithm, by Cauchy--Schwarz with $E_ph^2\le a_0$; each term linear in one $h_x$ contributes
$a_0\ln(1/a_0)$ times the density of its momentum at the origin ($1$ for $\bm p$, $\frac12$ for $\bm s$) times a
conditional angular average equal to $\frac12$, so
\begin{equation}
\alpha=1+\tfrac12+2\big(-\tfrac34\big)+\tfrac58+\tfrac58=\tfrac54 .
\end{equation}
The same counting reproduces the exact coefficient $\frac14$ of the mixed term, and quadrature down to $a_0=10^{-5}$
gives local slopes $1.192$, $1.223$, $1.239$, $1.245$, $1.248\to5/4$.

Second, the friction correction acquires an $a_0^2\ln^2(1/a_0)$ term. Its only source is the self-consistent part of the
mixed term, $c_{12}^{\rm SCBA}=\frac14(1+f)(f-a_0(1-f))$; the bounds above exclude one from $c_4$, confinement excludes one
from $\sigma_2^4$, and the remainder is
\begin{equation}
\begin{aligned}c_{12}-c_{12}^{\rm SCBA}={}&c_Ma_0-a_0^2\frac{\ln2}{4}\Big(\ln\frac1{a_0}-\gamma_E\Big)\\&+a_0^2\,\frac{\tau+\sigma_S}{2}+O(a_0^3\ln^2a_0),\end{aligned}
\end{equation}
with $\tau=\frac{\pi^2}{24}-\frac{\ln^22}{4}+\frac78-\frac54\ln2$ and $\sigma_S=\frac{17}{6}-4\ln2$. With
$a_0=AT^p$, the $T^{2p}\ln^2T$ coefficient of $\gamma_{\rm eff}/\gamma$ is exactly $\frac14\sigma_1^2\sigma_2^2p^2A^2$.

\subsection{No bound or quasibound state}
A localization-like enhancement would require a composite pole of two disorder-coupled lines. There is none. The
kinetic generator is accretive and the DC operator $L=\gamma(1+w_1)-\nabla\cdot\nu(\bm x)\nabla$ is self-adjoint with
$L\ge\gamma_{\min}$; from $G^2=-\partial_sG$ and ${\rm Tr}\,GG^{\rm T}={\rm Tr}\,G^2$, every same-frequency or equal-time
multi-line object is analytic off the single-line spectrum. In the conserving approximation (self-consistent Born plus
the same-frequency ladder) the Ward identity $\Gamma_{\rm vertex}=1+\partial_s\Sigma$ places the ladder pole exactly
at the edge of the dressed continuum. The bubble
$\int\frac{\dd\omega}{2\pi}G^R(\omega_+,\bm q_+)G^A(\omega_-,\bm q_-)=1/(a_++a_--\ii\Omega)$ with
$a_++a_-=2[\gamma+\nu q^2+\nu Q^2/4-\ii\Omega/2]$ shows that a would-be pair pole with diffusion constant $\nu/2$ is
simply the edge of the two-line continuum. Exact spectra of the random operator show a continuous edge with a Lifshitz
tail below the self-consistent edge and no isolated states.

\subsection{Numerical solution of the random Brinkman problem}
We solved the random Brinkman problem exactly in a periodic box (spectral Galerkin on the stream function, $256^2$ to
$512^2$ points, $\xi$ equal to four grid spacings) and compared disorder averages with the fourth-order coefficients.
The lognormal Darcy value $0.88247(14)$ agrees with Dykhne's $0.88254$ at lognormal width $s=0.5$; $c_4$ agrees within $1.1$
standard errors at $a_0=\infty$, 1, 0.3 and 0.1; $c_{12}$ within $0.2$ standard errors at $a_0=1$, 0.3 and 0.1, while the
self-consistent-Born value alone is excluded by 4 to 12 standard errors (12 at $a_0=1$); and $c_4^\nu$ within $1.0$--$1.6$ standard errors at
$a_0=0$, 0.1, 1 and 10, with the lognormal duality satisfied within $0.5$ standard errors.

\begin{figure}[t]
\includegraphics[width=\columnwidth]{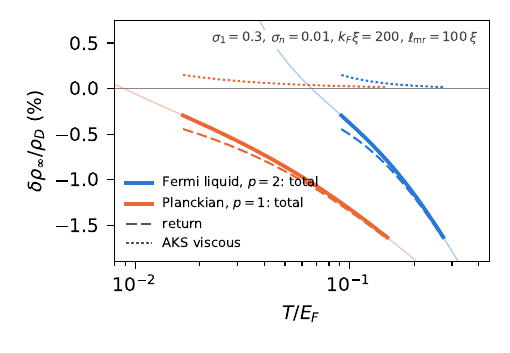}
\caption{\label{fig:aks}Bulk intercept: friction return and viscous density-landscape term for a Fermi liquid and a
Planckian fluid, inside the validity window. $\sigma_1=0.3$, $\sigma_n=0.01$, $k_F\xi=200$,
$\ell_{\rm mr}=100\,\xi$, $C_\eta=\alpha_P=1$; the thermal term is omitted.}
\end{figure}

\section{Density landscape}\label{app:aks}
Real samples may also modulate the electron density. Let $n=\bar n(1+w)$ with
$\langle w(\bm x)w(\bm x')\rangle=\sigma_n^2e^{-|\bm x-\bm x'|^2/2\xi^2}$, keep the viscosity uniform at this order, and
solve the Brinkman problem with continuity $\nabla\cdot(n\bm u)=0$ at DC. To second order the dissipation gives
\begin{equation}
\frac{\delta\rho}{\rho_D}=\frac{\sigma_n^2\nu}{\gamma\xi^2}+\frac{\sigma_n^2}{2}-\frac{\sigma_1^2}{2}f(a_0)
=\frac{\sigma_n^2}{2a_0}+\frac{\sigma_n^2}{2}-\frac{\sigma_1^2}{2}f(a_0).
\end{equation}
The first term is the viscous dissipation of the flow forced around the density inhomogeneities. It agrees with the
viscous term of Andreev, Kivelson and Spivak (AKS)~\cite{AKS2011},
$\rho=(1/2e^2)\langle T(\delta s_0)^2/\kappa+(\eta+\zeta)|\nabla(1/n_0)|^2\rangle$ with $s_0$ the entropy per particle:
for $\zeta=0$ and $\langle|\nabla w|^2\rangle=2\sigma_n^2/\xi^2$ that term is $\eta\sigma_n^2/(\bar n^2\xi^2)$, identical to
ours.

In a channel we estimate the landscape term locally, weighting the dissipation with the uniform-channel profile. It becomes
$P(x)\,\sigma_n^2h(x)/(2a_0)$ in $\delta\rho/\rho$, with
\begin{equation}
P(x)=\frac{x\,(x\cosh2x+2x-\tfrac32\sinh2x)}{2(x\cosh x-\sinh x)^2},
\end{equation}
which interpolates between $P=6/5=\langle u^2\rangle/\langle u\rangle^2$ for Poiseuille flow and $P=1$ for plug flow. In
narrow channels $Ph/(2a_0)\to1/(10r^2)$ with $r=\xi/W$, temperature independent, as the analyticity of the narrow-channel
resistivity (Appendix~\ref{app:channel}) requires. A wall layer for this term is not computed.

The thermal AKS term, $T\langle\delta s_0^2\rangle/(2\kappa)$, requires the energy sector, which the $m\le2$ hierarchy
does not contain. With kinetic $\kappa=\frac12Cv^2\tau_\kappa$ and $\eta=\bar np_Fv\tau_\eta/4$ its ratio to the viscous
term is $\alpha_{\rm th}=(2\pi^2/3)(T/E_F)^2\xi^2/(\ell_\kappa\ell_\eta)$, a relative $O((T/E_F)^2)$; at fixed $\gamma$ it
scales as $T^4$ in a Fermi liquid and as $T^3$ in a Planckian fluid. Its additivity to the momentum problem at $\gamma>0$
is not derived, and we omit it.

The bulk intercept is therefore a combination of $\{1,a_0^{-1},a_0\ln a_0,a_0\}$, plus $a_0^2$ (Fermi liquid) or $a_0^3$
(Planckian) if the thermal term is kept. These functions are independent (their Wronskians, $24/a^7$ and $192/a^6$, do
not vanish), so the return can be separated from both landscape terms in a fit against the measured viscosity; 
Appendix~\ref{app:widthscan} makes this explicit. Which term dominates depends on the amplitudes. For $\sigma_n\sim\sigma_1$ the
viscous landscape term exceeds the return by $1/a_0^2$; for weak density fluctuations away from neutrality,
$\sigma_n\ll\sigma_1$, the return can dominate (Fig.~\ref{fig:aks}).

\section{Channel}\label{app:channel}
\subsection{Uniform channel and the channel formula}
For a strip $|y|\le W/2$ with no-slip walls, the DC Brinkman equation $\nu u''-\gamma u+F=0$ with $D_\nu^2=\nu/\gamma$
is solved by
\begin{equation}
\begin{gathered}
u(y)=\frac F\gamma\Big[1-\frac{\cosh(y/D_\nu)}{\cosh(W/2D_\nu)}\Big],\\
\langle u\rangle=\frac F\gamma\,h(x),\qquad h(x)=1-\frac{\tanh x}{x},\qquad x=\frac W{2D_\nu},
\end{gathered}
\end{equation}
so $\rho_{\rm ch}=(m/ne^2)\gamma/h(x)$. The Ohmic limit is $\gamma(1+1/x)$, and the Poiseuille limit, from
$h=x^2/3+O(x^4)$, is $\rho\simeq12m\nu/ne^2W^2+(6/5)m\gamma/ne^2$. To first order the disorder corrections enter
through the two rates, $\delta\rho/\rho=(1-\eta)\delta\gamma/\gamma+\eta\,\delta\nu/\nu$ plus a wall term
[Eq.~(\ref{eq:channel})], with $\eta=\partial\ln\rho/\partial\ln\nu$ the viscous fraction of the resistance. The weights sum to one
(Euler), and $\eta$ equals the fraction of the dissipation that is viscous (Hellmann--Feynman).

\subsection{Wall layer}
Near a no-slip wall the stress-disorder correlations are cut off. From the clamped half-plane Green function, the image
contribution $-\sqrt{2/\pi}$ and the truncation of bulk correlations $+\sqrt{2/\pi}/15$ give
$c_w=-\frac{14}{15}\sqrt{2/\pi}$. The resulting wall term
$\Delta_w=-2c_w\sigma_2^2(\xi/W)\tanh^2x/h(x)$ is exactly a Navier slip of length
$\ell_s=c_w\sigma_2^2\xi=-0.745\,\sigma_2^2\xi$, an anti-slip, since $\dd\ln\rho/\dd\ell_s=-(2/W)\tanh^2x/h$. The identity
$\tanh^2x/h=1+2\eta$ makes the slip a rescaling of the two rates,
$\gamma^*=\gamma(1-2\ell_s/W)$ and $\nu^*=\nu(1+\sigma_2^2/2-6\ell_s/W)$. For other correlator shapes
$\ell_s=-(28/15\pi)\int_0^\infty\langle w_2w_2\rangle(\rho)\,\dd\rho$.

\subsection{Wide and narrow channels}
In wide channels ($x\gg1$) the $x^0$ term is the bulk $\gamma_{\rm eff}/\gamma$ exactly. The $1/x$ term contains the friction
wall correction $-\sigma_1^2a_0\chi/x$ with $\chi=-0.472624357$, a stress wall term
$\sigma_2^2(a_0/x)[-3.5\ln(D_\nu/\xi)+3.52532]$, and, at fourth order, the bulk logarithm
$a_0\ln(1/a_0)(9\sigma_2^4+\sigma_1^2\sigma_2^2)/16$. The leading return term, $-a_0\ln(1/a_0)\propto-T^p\ln(1/T)$, cannot
be absorbed into the bare parameters: the Wronskian of $\{1,T^{-p/2},T^p\ln T\}$ is
$-p^2T^{p/2-3}(3p\ln T+5)/4\neq0$.

In narrow channels ($x\ll1$), with $r=\xi/W$ and $\ell=\ln(W/\xi)$,
\begin{equation}
\begin{aligned}\frac{\delta\rho}{\rho}={}&\frac{\sigma_2^2}{2}+\frac{28}{5}\sqrt{\frac2\pi}\sigma_2^2r+\sigma_2^2r^2[B_2-24\ell]\\&-\frac{\sigma_2^2}{5}x^2\Big[1+\frac{112}{15}\sqrt{\frac2\pi}r\Big]+\dots,\end{aligned}
\end{equation}
with $B_2=38.4599$ and an $x^2r^2$ coefficient $\frac25\ell+9.1033$; these constants were reproduced by an exact strip
calculation that does not use the expansion. Writing $\rho=a\nu/W^2+b\gamma+c\gamma^2W^2/\nu$, equivalently $\rho=(\nu/W^2)(a+4bx^2+16cx^4)$, the uniform channel
has $(a,b,c)=(12,6/5,-1/700)$. In the notation of Sec.~\ref{sec:chanresp}, $\rho=\rho_P+\frac65\rho_D-\frac{3}{175}(\rho_D^2/\rho_P)(1+\Delta)$, so that
$\Delta=\delta R/R$ exactly. Every correction of order $\sigma^2$ or $r$ only rescales $\gamma$ and $\nu$, which
leaves the combination $R=ac/b^2=-1/84$ unchanged. To find what does change it, write the disorder correction as
$\delta\rho/\rho=d_0+d_1x^2+d_2x^4$ and linearize $R$:
\begin{equation}
\frac{\delta R}{R}=-215\,d_1-525\,d_2 ;
\label{eq:SdR}
\end{equation}
$d_0$, a rescaling of $\nu$, drops out. Only the $r^2$ parts of $d_1$ and $d_2$ contribute.

\emph{Stress disorder.} Its $r^2$ coefficients are $d_1=\sigma_2^2r^2(\frac25\ell+9.1033)$ and
$d_2=\sigma_2^2r^2(-\frac{132}{175}\ell-1.5091)$. The log coefficients are exact: they are the Taylor coefficients of the
stress logarithm in the strip, known in closed form in terms of the uniform-channel profile,
\begin{equation}
\begin{aligned}
\ell(x)&=\frac{4x^2}{h(x)}\Big[-\frac{2\tanh x}{x}+\frac{\sinh 2x/2x-1}{4\cosh^2x}\Big]\\
&=-24+\tfrac25x^2-\tfrac{132}{175}x^4+\dots ,
\end{aligned}
\end{equation}
whose $x^0$ term is the wall logarithm absorbed into $\nu$; the constants combine strip
integrals evaluated by quadrature and are reproduced by the calculation exact in $W/D_\nu$ to $10^{-5}$ and $3\times10^{-4}$.
Inserting them in Eq.~(\ref{eq:SdR}),
\begin{equation}
\begin{aligned}
\frac{\delta R}{R}\Big|_{\sigma_2^2}&=\sigma_2^2r^2\big[(-86+396)\,\ell-1957.21+792.28\big]\\
&=310\,\sigma_2^2r^2\big[\ell-3.758\big],
\end{aligned}
\end{equation}
where the fit precision of $d_2$ gives $\pm6\times10^{-4}$ on $3.758$.

\emph{Friction disorder.} In a narrow channel the friction return is cut off at $W$ and enters only the $\gamma^2$ term,
$\delta c=-\frac35\sigma_1^2r^2L_1$, so $\delta R/R=\delta c/c=420\,\sigma_1^2r^2L_1$. Physically, the wall-cut return
lowers the friction by $(\delta\gamma/\gamma)_W=-\sigma_1^2a_0L_1$ (the bulk $\ln(1/a_0)$ becomes $2L_1$); since this is
itself $\propto\gamma^2/\nu$ it feeds the curvature term, and with $a_0=6r^2\rho_D/\rho_P$,
$\Delta=-70(\rho_P/\rho_D)(\delta\gamma/\gamma)_W=420\,\sigma_1^2r^2L_1$, where $70=(6/5)/(3/175)$ is the ratio of the
uniform-channel Drude and curvature coefficients. The logarithm
$L_1=\ln(W/\rho_c)+\frac12+\langle h_s\rangle$ follows from the exact clamped-strip Green function. Here
$\rho_c=e^{\langle\ln\rho\rangle}$ is the mean log-distance of the correlator, $\rho_c=\sqrt2\,e^{-\gamma_E/2}\xi$ for the
Gaussian, and $\langle h_s\rangle=-2.05867$ is a pure strip-geometry number with Poiseuille weight, independent of the
correlator and evaluated by quadrature of a closed-form integrand. Hence $L_1=\ell+c_L$ with
\begin{equation}
c_L=\tfrac12\gamma_E-\tfrac12\ln2+\tfrac12+\langle h_s\rangle=-1.6166 .
\end{equation}
Together,
\begin{equation}
\begin{aligned}\Delta=\frac{\delta R}{R}={}&r^2\big\{420\sigma_1^2[\ell+c_L]+310\sigma_2^2[\ell-3.758]\\&-88.73\,\sigma_2^4\big\},\end{aligned}
\label{eq:SDelta}
\end{equation}

\emph{Slip.} A Navier slip $\lambda=\ell_s/W$ gives exactly
\begin{equation}
\begin{gathered}
R(\lambda)=-\frac{1+20\lambda}{84(1+10\lambda+30\lambda^2)^2},\\
\frac{\delta R}{R}=-160\lambda^2+O(\lambda^3),
\end{gathered}
\end{equation}
so the first order in slip cancels. With the disorder slip $\lambda=\ell_s/W=c_w\sigma_2^2r$ the square gives $\Delta=-160\ell_s^2/W^2=
-160c_w^2\sigma_2^4r^2=-(12544/45\pi)\,\sigma_2^4r^2=-88.73\,\sigma_2^4r^2$, the last term of Eq.~(\ref{eq:SDelta}). The other $O(\sigma^4r^2)$ strip terms
are not computed and are estimated to be a relative $O(\sigma^2)$ correction.

\emph{Size.} For $\sigma_1=\sigma_2=0.3$, Eq.~(\ref{eq:SDelta}) gives $\Delta=0.075$, $0.036$ and $0.014$ at $W/\xi=20$, $50$
and $100$. The curvature term it multiplies is $(3/175)(\rho_D/\rho_P)^2\rho$ to leading order, with $\rho_D/\rho_P=x^2/3$:
$1.1\times10^{-4}$ of $\rho$ at $x=0.5$ and $1.4\times10^{-3}$ at $x=1$. Detecting $\Delta$ therefore requires the
$T^p$ coefficient of $\rho(T)$ to a few percent, a shift of $\rho$ of order $10^{-5}$--$10^{-4}$.

\subsection{Analyticity of the narrow-channel resistivity}
In narrow channels the walls cut the return integral, and the resistivity is analytic in $x^2\propto T^p$ to all orders
in the disorder. The argument has four steps. (i) Korn's inequality: for no-slip, divergence-free flows,
$\frac12\int\nu|\bm S|^2\ge\nu_{\min}\int|\nabla\bm u|^2$. (ii) Poincar\'e's inequality across the strip,
$\int_0^W|\partial_y\bm u|^2\ge(\pi/W)^2\int_0^W|\bm u|^2$, so the viscous operator is bounded below by
$\mu=\nu_{\min}\pi^2/W^2$ uniformly in the strip length. (iii) With the friction weight $0\le m\le m_{\max}$, the operator
$A+\gamma M$ is invertible off $(-\infty,-\mu/m_{\max}]$, and the flux is a Stieltjes function of $\gamma$, so
$\rho=1/q$ is holomorphic there~\cite{Kato1995}. (iv) Writing $A=\nu A_1$, $\rho=\nu R(x^2)$ with Taylor radius
$x_c^2=(\pi^2/4)(\nu_{\min}/\nu)/m_{\max}$; with $\nu\propto T^{-p}$, $\rho(T)$ contains integer powers of $T^p$ only, and for a general $\Gamma_2(T)$ it is a power
series in $x^2\propto\gamma\Gamma_2(T)$, with no $\ln a_0$. For
Gaussian disorder the statement holds order by order. The uniform-channel flux checks the bound: by Mittag-Leffler,
$h/x^2=\sum_n\frac{8}{(2n+1)^2\pi^2}[x^2+(2n+1)^2\pi^2/4]^{-1}$, whose first pole sits exactly at $x^2=-\pi^2/4$. In the
opposite limit $W\to\infty$ the gap closes and the bulk $e^{a_0}E_1(a_0)$ contains $\ln a_0$.

\subsection{Temperature crossover at fixed width}
At fixed $W$, $x^2=a_0/2r^2$ and the channel correction groups as
\begin{equation}
\frac{\delta\rho}{\rho}=\underbrace{(1-\eta)\frac{\delta\gamma}{\gamma}+\eta\frac{\delta\nu}{\nu}}_{\text{friction + viscosity}}
+\underbrace{\Delta_w+P\,\frac{\sigma_n^2h}{2a_0}}_{\text{wall + landscape}} .
\end{equation}
Only the first group carries $\ln T$, through $f(a)=a[\ln(1/a)-\gamma_E]+O(a^2\ln a)$ with $\ln(1/a_0)=p\ln(T_\xi/T)$; the
wall and landscape factors are even power series in $x$, and $\eta(x)=1-\frac25x^2+\dots$ is analytic in $x^2$. In wide
channels $\eta\to1/2x$ suppresses the viscosity logarithm to
$(\sigma_2^2/2)(r/\sqrt2)\sqrt{a_0}\ln(1/a_0)\propto T^{p/2}\ln(1/T)$. In narrow channels
$a_0\ln(1/a_0)=2r^2x^2[2\ln(W/\xi)-\ln2-\ln x^2]$: the first term is a logarithm of the width, and the $\ln x^2$ term is
removed by the walls. Eq.~(\ref{eq:channel}), which uses the bulk $f$ at every $x$, retains this term, but its size is bounded
by $\frac{\sigma^2}{2}\,2r^2\max_{x<1}x^2|\ln x^2|=\sigma^2r^2/e$, with $\sigma=\max(\sigma_1,\sigma_2)$: for $W=20\,\xi$ and $\sigma_{1,2}=0.3$ it
stays below $10^{-4}$ of $\rho$. The friction-plus-viscosity group changes sign where the friction return
overtakes the viscous rescaling [Fig.~\ref{fig:widthscan}(a)].

\section{Temperature dependence and the width-scan protocol}\label{app:widthscan}
\subsection{Temperature dependence}
\begin{figure}[t]
\includegraphics[width=0.85\columnwidth]{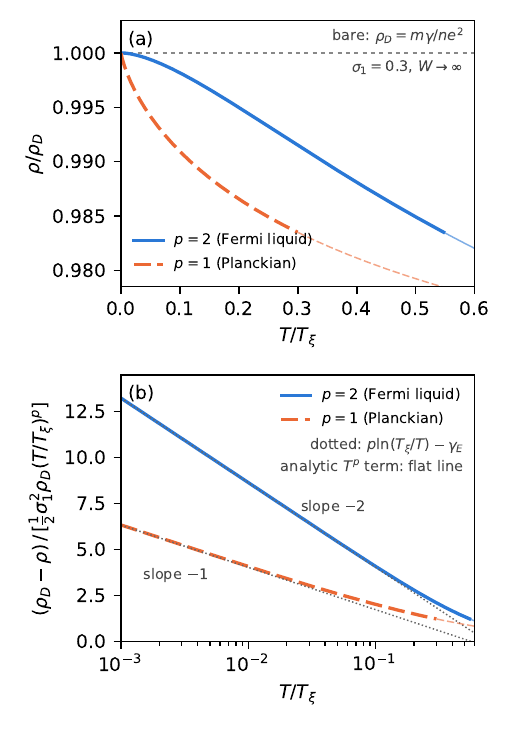}
\caption{\label{fig:rhoT}Bulk resistivity. (a) $\rho/\rho_D$ against $T/T_\xi$ for $\sigma_1=0.3$. (b) The correction
divided by its analytic prefactor, $(\rho_D-\rho)/[\frac12\sigma_1^2\rho_D(T/T_\xi)^p]$: a straight line in $\ln T$
with slope $-p$, where an analytic $T^p$ term would be flat.}
\end{figure}

With $\gamma$, $\sigma_1$ and $\xi$ temperature independent and $\Gamma_2\propto T^p$ ($p=2$ for a Fermi liquid,
$p=1$ for a Planckian fluid), $a_0=2\gamma\Gamma_2\xi^2/v^2=(T/T_\xi)^p$, and the bulk resistivity is
$\rho/\rho_D=1-(\sigma_1^2/2)f(a_0)$. Since
$e^{a_0}E_1(a_0)=\ln(1/a_0)-\gamma_E+a_0(1-\gamma_E-\ln a_0)+o(a_0)$, the correction divided by its analytic prefactor
is a straight line in $\ln T$ with slope $-p$, and it vanishes non-analytically as $T\to0$ (Fig.~\ref{fig:rhoT}). Every
disorder quantity depends on temperature only through $(\gamma,\Gamma_2)$, via $x^2\propto\gamma\Gamma_2$, $a_0=2r^2x^2$
and $\ell_\nu=v/2\Gamma_2$. At fixed $\gamma$ the corrections are therefore the same functions of $a_0$ for the two fluids;
only the temperature axis differs, and the Planckian window spans about two decades in $T$.

How large may the disorder be? Comparing the Born channel resistivity with its Gaussian fourth-order completion along
channel curves at fixed width inside the validity window ($\sigma_1=\sigma_2$, $W=20\,\xi$ and $50\,\xi$, $p=1$ and 2), the
largest deviation stays below $10\%$ of the Born correction for $\sigma_2\lesssim0.28$ and below $20\%$ for
$\sigma_2\lesssim0.40$; at $\sigma_{1,2}=0.3$, the value used in the figures, it is $10.5$--$10.8\%$. In the Poiseuille limit,
which lies outside the window, the ratio tends analytically to $(33/4-10\ln2)\sigma_2^2$. Beyond Born order the result depends on the ensemble: for a lognormal stress
rate the Stokes duality gives $c_4^\nu(0)=-1/8$ instead of the Gaussian $0.659$.

\subsection{Window in measurable lengths}
With $\ell_g=v/\gamma$ and $\ell_\Gamma=v/\Gamma_2$, $a_0=2\xi^2/(\ell_g\ell_\Gamma)$ and
$D_\nu=\frac12\sqrt{\ell_g\ell_\Gamma}$. The window $\ell_\nu\le0.3\,\xi$, $a_0\le0.3$ spans
$a_{\max}/a_{\min}=(9/100)\,\ell_g/\xi$: one decade requires $\ell_g\ge111\,\xi$, and any window requires
$\ell_g\ge(100/9)\,\xi$. Wide channels, $x\ge4$, require $W\ge4\sqrt2\,\xi/\sqrt{a_0}$.

\subsection{The width scan}
At fixed $T$ a family of channels of different widths separates the two rates. Narrow channels give
$\rho_N=12\nu^*/W^2+\frac65\gamma^*$: the slope of $\rho_N$ against $1/W^2$ is $S_N=12(m/ne^2)\nu_N$ with
$\nu_N=\nu(1+\sigma_2^2/2)$, the bare $\nu(T)$ times a temperature-independent factor, and the disorder slip appears only
in odd powers of $1/W$. Wide channels, extrapolated to $\xi/W\to0$, give the bulk intercept $\rho_\infty$ exactly. The
return parameter is then
\begin{equation}
a_N=\frac{\xi^2\gamma}{2\nu_N}=\frac{6\xi^2\rho_D}{S_N},\qquad \rho_D=\frac{m\gamma}{ne^2},
\end{equation}
so the measured $1/S_N(T)$ fixes $a_N$ up to the constant $6\xi^2\rho_D$. Extrapolating wide channels with the uniform-channel
shape $\gamma^*/h(W/2D^*)$ recovers the intercept with a bias of at most $1.1\%$ of the signal.

Without a density landscape, $\rho_\infty(T\to0)=\rho_D$ and $y=[1-\rho_\infty/\rho_\infty(0)]/a_N$ is, at leading
logarithm, a straight line in $\ln a_N$ whose slope and intercept return $\xi^2(1+\sigma_2^2/2)$ and $\sigma_1^2$. A
landscape changes this in one respect. In narrow channels it adds $\frac65(m/ne^2)\sigma_n^2\nu/\xi^2$, independent of
$W$, so the slope $S_N$ and hence $a_N$ are unaffected. In wide channels
\begin{equation}
\frac{\rho_\infty}{\rho_D}=1+\frac{\sigma_n^2}{2}+\frac{\sigma_n^2}{2a_0}-\frac{\sigma_1^2}{2}f(a_0),\qquad a_0=\kappa'z,
\end{equation}
with $z=1/S_N$ measured and $\kappa'=6\xi^2\rho_D(1+\sigma_2^2/2)$ unknown. Because $\rho_\infty$ now grows as $T\to0$,
$\rho_\infty(0)$ is no longer a usable reference; instead one fits. At leading logarithm
$\rho_\infty=A+B/z+Cz\ln z+Dz$, with
\begin{equation}
\begin{gathered}A=\rho_D\Big(1+\frac{\sigma_n^2}{2}\Big),\qquad B=\frac{\rho_D\sigma_n^2}{2\kappa'},\\ C=\frac{\sigma_1^2\rho_D\kappa'}{2},\qquad D=C\,(\ln\kappa'+\gamma_E),\end{gathered}
\end{equation}
and the coefficients invert uniquely:
\begin{equation}
\begin{gathered}\kappa'=e^{D/C-\gamma_E},\qquad \rho_D=A-B\kappa',\\ \sigma_n^2=\frac{2B\kappa'}{\rho_D},\qquad \sigma_1^2=\frac{2C}{\rho_D\kappa'} .\end{gathered}
\end{equation}
The four functions are independent, with Wronskian $W[1,1/z,z\ln z,z]=4/z^5$, so the protocol returns $\sigma_n^2$ as
well as $\sigma_1^2$ and $\xi$. The quantity plotted in Fig.~\ref{fig:widthscan}(c) is $y=[1-\rho_\infty/\bar\rho]/a_N$ with
$\bar\rho=\rho_D(1+\sigma_n^2/2)$ from the fit. It crosses zero where $a_0f(a_0)=\sigma_n^2/\sigma_1^2$, at
$a_0\simeq0.018$ for $\sigma_n=0.01$ and $\sigma_1=0.3$; above that point the return dominates. Applied to synthetic
channel data that include the landscape, the uniform-channel extrapolation stays within $1.1\%$ of the intercept signal,
and a fit of the extrapolated intercepts with the exact bulk form recovers $(\rho_D,\kappa',\sigma_1^2,\sigma_n^2)$ to
better than $0.2\%$ from starting values $30\%$ off. This second test is noiseless and uses the same functional form as
the data, so it checks the conditioning of the fit rather than the physics.

\section{Validity window and higher harmonics}\label{app:window}
For any tower of harmonics the Born results depend on it only through the wavevector-dependent viscosity $\nu(k)$:
$\gamma_R/\gamma=1-\frac{\sigma_1^2}{2}\langle[1+k^2D_\nu^2\nu(k)/\nu]^{-1}\rangle$ and
$\delta\nu/\nu=\frac{\sigma_2^2}{2}\langle\nu(k)/\nu+[\nu/\nu(k)+k^2D_\nu^2]^{-1}\rangle$. Keeping $m=\pm3$, with
$\varepsilon=\gamma/\Gamma_3=\ell_\nu^2/D_\nu^2$,
\begin{equation}
\begin{gathered}
\frac{\gamma_R}{\gamma}=1-\frac{\sigma_1^2}{2}\,\frac{\varepsilon+f(a_0/(1+\varepsilon))}{1+\varepsilon},\\
\frac{\delta\nu}{\nu}=\frac{\sigma_2^2}{2}\Big[f\Big(\frac{a_0}{\varepsilon}\Big)+f\Big(\frac{a_0}{1+\varepsilon}\Big)\Big],
\end{gathered}
\end{equation}
which adds the local terms $-(\sigma_1^2/2)\gamma/\Gamma_3$ and $-\sigma_2^2\ell_\nu^2/\xi^2$; $m\ge4$ enters at relative
$O(\ell_4^2/\xi^2)$. The corrections are therefore controlled for $\gamma/2\Gamma_3\ll a_0\ll1$, i.e.
$\ell_\nu\ll\xi\ll D_\nu$; terms built on the stress return need $\ell_\nu D_\nu\ll\xi^2[\ln(1/a_0)/4]^{1/2}$. With
$\nu\propto T^{-p}$ and $\ell_\nu\propto T^{-q}$ the window is $T_0(\ell_*/\xi)^{1/q}<T<T_0(D_*/\xi)^{2/p}$, with
$q=(p+p_3)/2$ in the kinetic model. In channels with $m=\pm3$ kept, $\nu_{\rm eff}=\nu(1+\gamma/\Gamma_3)$ and no new
boundary layer appears; with the minimal wall condition the $m=3$ layer adds a Navier slip $\sigma_2^2\ell_\nu/4$, so
the total disorder slip is $\ell_s=\sigma_2^2(c_w\xi+\ell_\nu/4)$.

A strictly positive shot-noise ensemble with exactly Gaussian covariance exists for $\sigma^2\le1/(2N_\xi)$, and the
Born results do not depend on the ensemble. Beyond Born order a positive, non-Gaussian ensemble adds a third-cumulant
term of order $\sigma^3$ times the skewness; the fourth-order results above refer to the Gaussian ensemble. We take
$\sigma_1$ temperature independent and $\sigma_2(T)=\sigma_\lambda|\partial\ln\Gamma_2/\partial\ln\lambda|_T$, where
$\lambda$ is the fixed-density parameter that modulates $\Gamma_2$; a separable $\Gamma_2=\Phi(\lambda)T^p$ gives a
constant $\sigma_2$. The two disorders are taken as independent; a cross-correlation would add an $O(q)$ mixed
self-energy, not treated here.


\begin{thebibliography}{34}%
\makeatletter
\providecommand \@ifxundefined [1]{%
 \@ifx{#1\undefined}
}%
\providecommand \@ifnum [1]{%
 \ifnum #1\expandafter \@firstoftwo
 \else \expandafter \@secondoftwo
 \fi
}%
\providecommand \@ifx [1]{%
 \ifx #1\expandafter \@firstoftwo
 \else \expandafter \@secondoftwo
 \fi
}%
\providecommand \natexlab [1]{#1}%
\providecommand \enquote  [1]{``#1''}%
\providecommand \bibnamefont  [1]{#1}%
\providecommand \bibfnamefont [1]{#1}%
\providecommand \citenamefont [1]{#1}%
\providecommand \href@noop [0]{\@secondoftwo}%
\providecommand \href [0]{\begingroup \@sanitize@url \@href}%
\providecommand \@href[1]{\@@startlink{#1}\@@href}%
\providecommand \@@href[1]{\endgroup#1\@@endlink}%
\providecommand \@sanitize@url [0]{\catcode `\\12\catcode `\$12\catcode
  `\&12\catcode `\#12\catcode `\^12\catcode `\_12\catcode `\%12\relax}%
\providecommand \@@startlink[1]{}%
\providecommand \@@endlink[0]{}%
\providecommand \url  [0]{\begingroup\@sanitize@url \@url }%
\providecommand \@url [1]{\endgroup\@href {#1}{\urlprefix }}%
\providecommand \urlprefix  [0]{URL }%
\providecommand \Eprint [0]{\href }%
\providecommand \doibase [0]{https://doi.org/}%
\providecommand \selectlanguage [0]{\@gobble}%
\providecommand \bibinfo  [0]{\@secondoftwo}%
\providecommand \bibfield  [0]{\@secondoftwo}%
\providecommand \translation [1]{[#1]}%
\providecommand \BibitemOpen [0]{}%
\providecommand \bibitemStop [0]{}%
\providecommand \bibitemNoStop [0]{.\EOS\space}%
\providecommand \EOS [0]{\spacefactor3000\relax}%
\providecommand \BibitemShut  [1]{\csname bibitem#1\endcsname}%
\let\auto@bib@innerbib\@empty
\bibitem [{\citenamefont {Gor'kov}\ \emph {et~al.}(1979)\citenamefont
  {Gor'kov}, \citenamefont {Larkin},\ and\ \citenamefont
  {Khmel'nitskii}}]{GLK1979}%
  \BibitemOpen
  \bibfield  {author} {\bibinfo {author} {\bibfnamefont {L.~P.}\ \bibnamefont
  {Gor'kov}}, \bibinfo {author} {\bibfnamefont {A.~I.}\ \bibnamefont
  {Larkin}},\ and\ \bibinfo {author} {\bibfnamefont {D.~E.}\ \bibnamefont
  {Khmel'nitskii}},\ }\bibfield  {title} {\bibinfo {title} {Particle
  conductivity in a two-dimensional random potential},\ }\href@noop {}
  {\bibfield  {journal} {\bibinfo  {journal} {JETP Lett.}\ }\textbf {\bibinfo
  {volume} {30}},\ \bibinfo {pages} {228} (\bibinfo {year} {1979})}\BibitemShut
  {NoStop}%
\bibitem [{\citenamefont {Lee}\ and\ \citenamefont
  {Ramakrishnan}(1985)}]{LeeRamakrishnan1985}%
  \BibitemOpen
  \bibfield  {author} {\bibinfo {author} {\bibfnamefont {P.~A.}\ \bibnamefont
  {Lee}}\ and\ \bibinfo {author} {\bibfnamefont {T.~V.}\ \bibnamefont
  {Ramakrishnan}},\ }\bibfield  {title} {\bibinfo {title} {Disordered
  electronic systems},\ }\href@noop {} {\bibfield  {journal} {\bibinfo
  {journal} {Rev. Mod. Phys.}\ }\textbf {\bibinfo {volume} {57}},\ \bibinfo
  {pages} {287} (\bibinfo {year} {1985})}\BibitemShut {NoStop}%
\bibitem [{\citenamefont {Bergmann}(1984)}]{Bergmann1984}%
  \BibitemOpen
  \bibfield  {author} {\bibinfo {author} {\bibfnamefont {G.}~\bibnamefont
  {Bergmann}},\ }\bibfield  {title} {\bibinfo {title} {Weak localization in
  thin films: a time-of-flight experiment with conduction electrons},\
  }\href@noop {} {\bibfield  {journal} {\bibinfo  {journal} {Phys. Rep.}\
  }\textbf {\bibinfo {volume} {107}},\ \bibinfo {pages} {1} (\bibinfo {year}
  {1984})}\BibitemShut {NoStop}%
\bibitem [{\citenamefont {Altshuler}\ \emph {et~al.}(1982)\citenamefont
  {Altshuler}, \citenamefont {Aronov},\ and\ \citenamefont
  {Khmelnitskii}}]{AAK1982}%
  \BibitemOpen
  \bibfield  {author} {\bibinfo {author} {\bibfnamefont {B.~L.}\ \bibnamefont
  {Altshuler}}, \bibinfo {author} {\bibfnamefont {A.~G.}\ \bibnamefont
  {Aronov}},\ and\ \bibinfo {author} {\bibfnamefont {D.~E.}\ \bibnamefont
  {Khmelnitskii}},\ }\bibfield  {title} {\bibinfo {title} {Effects of
  electron-electron collisions with small energy transfers on quantum
  localisation},\ }\href@noop {} {\bibfield  {journal} {\bibinfo  {journal} {J.
  Phys. C}\ }\textbf {\bibinfo {volume} {15}},\ \bibinfo {pages} {7367}
  (\bibinfo {year} {1982})}\BibitemShut {NoStop}%
\bibitem [{\citenamefont {Hikami}\ \emph {et~al.}(1980)\citenamefont {Hikami},
  \citenamefont {Larkin},\ and\ \citenamefont {Nagaoka}}]{HLN1980}%
  \BibitemOpen
  \bibfield  {author} {\bibinfo {author} {\bibfnamefont {S.}~\bibnamefont
  {Hikami}}, \bibinfo {author} {\bibfnamefont {A.~I.}\ \bibnamefont {Larkin}},\
  and\ \bibinfo {author} {\bibfnamefont {Y.}~\bibnamefont {Nagaoka}},\
  }\bibfield  {title} {\bibinfo {title} {Spin-orbit interaction and
  magnetoresistance in the two dimensional random system},\ }\href@noop {}
  {\bibfield  {journal} {\bibinfo  {journal} {Prog. Theor. Phys.}\ }\textbf
  {\bibinfo {volume} {63}},\ \bibinfo {pages} {707} (\bibinfo {year}
  {1980})}\BibitemShut {NoStop}%
\bibitem [{\citenamefont {Hershfield}\ and\ \citenamefont
  {Ambegaokar}(1986)}]{HA1986}%
  \BibitemOpen
  \bibfield  {author} {\bibinfo {author} {\bibfnamefont {S.}~\bibnamefont
  {Hershfield}}\ and\ \bibinfo {author} {\bibfnamefont {V.}~\bibnamefont
  {Ambegaokar}},\ }\bibfield  {title} {\bibinfo {title} {Transport equation for
  weakly localized electrons},\ }\href@noop {} {\bibfield  {journal} {\bibinfo
  {journal} {Phys. Rev. B}\ }\textbf {\bibinfo {volume} {34}},\ \bibinfo
  {pages} {2147} (\bibinfo {year} {1986})}\BibitemShut {NoStop}%
\bibitem [{\citenamefont {Narozhny}\ \emph {et~al.}(2002)\citenamefont
  {Narozhny}, \citenamefont {Zala},\ and\ \citenamefont {Aleiner}}]{NZA2002}%
  \BibitemOpen
  \bibfield  {author} {\bibinfo {author} {\bibfnamefont {B.~N.}\ \bibnamefont
  {Narozhny}}, \bibinfo {author} {\bibfnamefont {G.}~\bibnamefont {Zala}},\
  and\ \bibinfo {author} {\bibfnamefont {I.~L.}\ \bibnamefont {Aleiner}},\
  }\bibfield  {title} {\bibinfo {title} {Interaction corrections at
  intermediate temperatures: Dephasing time},\ }\href
  {https://doi.org/10.1103/PhysRevB.65.180202} {\bibfield  {journal} {\bibinfo
  {journal} {Phys. Rev. B}\ }\textbf {\bibinfo {volume} {65}},\ \bibinfo
  {pages} {180202} (\bibinfo {year} {2002})}\BibitemShut {NoStop}%
\bibitem [{\citenamefont {Ledwith}\ \emph {et~al.}(2019)\citenamefont
  {Ledwith}, \citenamefont {Guo},\ and\ \citenamefont {Levitov}}]{LGL2019}%
  \BibitemOpen
  \bibfield  {author} {\bibinfo {author} {\bibfnamefont {P.~J.}\ \bibnamefont
  {Ledwith}}, \bibinfo {author} {\bibfnamefont {H.}~\bibnamefont {Guo}},\ and\
  \bibinfo {author} {\bibfnamefont {L.}~\bibnamefont {Levitov}},\ }\bibfield
  {title} {\bibinfo {title} {The hierarchy of excitation lifetimes in
  two-dimensional {F}ermi gases},\ }\href@noop {} {\bibfield  {journal}
  {\bibinfo  {journal} {Ann. Phys.}\ }\textbf {\bibinfo {volume} {411}},\
  \bibinfo {pages} {167913} (\bibinfo {year} {2019})},\ \bibinfo {note}
  {arXiv:1905.03751}\BibitemShut {NoStop}%
\bibitem [{\citenamefont {Gurzhi}(1963)}]{Gurzhi1963}%
  \BibitemOpen
  \bibfield  {author} {\bibinfo {author} {\bibfnamefont {R.~N.}\ \bibnamefont
  {Gurzhi}},\ }\bibfield  {title} {\bibinfo {title} {Minimum of resistance in
  impurity-free conductors},\ }\href@noop {} {\bibfield  {journal} {\bibinfo
  {journal} {Sov. Phys. JETP}\ }\textbf {\bibinfo {volume} {17}},\ \bibinfo
  {pages} {521} (\bibinfo {year} {1963})}\BibitemShut {NoStop}%
\bibitem [{\citenamefont {Gurzhi}(1968)}]{Gurzhi1968}%
  \BibitemOpen
  \bibfield  {author} {\bibinfo {author} {\bibfnamefont {R.~N.}\ \bibnamefont
  {Gurzhi}},\ }\bibfield  {title} {\bibinfo {title} {Hydrodynamic effects in
  solids at low temperature},\ }\href@noop {} {\bibfield  {journal} {\bibinfo
  {journal} {Sov. Phys. Usp.}\ }\textbf {\bibinfo {volume} {11}},\ \bibinfo
  {pages} {255} (\bibinfo {year} {1968})}\BibitemShut {NoStop}%
\bibitem [{\citenamefont {Molenkamp}\ and\ \citenamefont
  {de~Jong}(1994)}]{Molenkamp1994}%
  \BibitemOpen
  \bibfield  {author} {\bibinfo {author} {\bibfnamefont {L.~W.}\ \bibnamefont
  {Molenkamp}}\ and\ \bibinfo {author} {\bibfnamefont {M.~J.~M.}\ \bibnamefont
  {de~Jong}},\ }\bibfield  {title} {\bibinfo {title} {Observation of knudsen
  and gurzhi transport regimes in a two-dimensional wire},\ }\href@noop {}
  {\bibfield  {journal} {\bibinfo  {journal} {Solid-State Electronics}\
  }\textbf {\bibinfo {volume} {37}},\ \bibinfo {pages} {551} (\bibinfo {year}
  {1994})}\BibitemShut {NoStop}%
\bibitem [{\citenamefont {de~Jong}\ and\ \citenamefont
  {Molenkamp}(1995)}]{deJong1995}%
  \BibitemOpen
  \bibfield  {author} {\bibinfo {author} {\bibfnamefont {M.~J.~M.}\
  \bibnamefont {de~Jong}}\ and\ \bibinfo {author} {\bibfnamefont {L.~W.}\
  \bibnamefont {Molenkamp}},\ }\bibfield  {title} {\bibinfo {title}
  {Hydrodynamic electron flow in high-mobility wires},\ }\href@noop {}
  {\bibfield  {journal} {\bibinfo  {journal} {Phys. Rev. B}\ }\textbf {\bibinfo
  {volume} {51}},\ \bibinfo {pages} {13389} (\bibinfo {year}
  {1995})}\BibitemShut {NoStop}%
\bibitem [{\citenamefont {Narozhny}\ \emph {et~al.}(2017)\citenamefont
  {Narozhny}, \citenamefont {Gornyi}, \citenamefont {Mirlin},\ and\
  \citenamefont {Schmalian}}]{Narozhny2017}%
  \BibitemOpen
  \bibfield  {author} {\bibinfo {author} {\bibfnamefont {B.~N.}\ \bibnamefont
  {Narozhny}}, \bibinfo {author} {\bibfnamefont {I.~V.}\ \bibnamefont
  {Gornyi}}, \bibinfo {author} {\bibfnamefont {A.~D.}\ \bibnamefont {Mirlin}},\
  and\ \bibinfo {author} {\bibfnamefont {J.}~\bibnamefont {Schmalian}},\
  }\bibfield  {title} {\bibinfo {title} {Hydrodynamic approach to electronic
  transport in graphene},\ }\href@noop {} {\bibfield  {journal} {\bibinfo
  {journal} {Annalen der Physik}\ }\textbf {\bibinfo {volume} {529}},\ \bibinfo
  {pages} {1700043} (\bibinfo {year} {2017})}\BibitemShut {NoStop}%
\bibitem [{\citenamefont {Lucas}\ and\ \citenamefont
  {Fong}(2018)}]{LucasFong2018}%
  \BibitemOpen
  \bibfield  {author} {\bibinfo {author} {\bibfnamefont {A.}~\bibnamefont
  {Lucas}}\ and\ \bibinfo {author} {\bibfnamefont {K.~C.}\ \bibnamefont
  {Fong}},\ }\bibfield  {title} {\bibinfo {title} {Hydrodynamics of electrons
  in graphene},\ }\href@noop {} {\bibfield  {journal} {\bibinfo  {journal} {J.
  Phys.: Condens. Matter}\ }\textbf {\bibinfo {volume} {30}},\ \bibinfo {pages}
  {053001} (\bibinfo {year} {2018})}\BibitemShut {NoStop}%
\bibitem [{\citenamefont {Fritz}\ and\ \citenamefont
  {Scaffidi}(2024)}]{FritzScaffidi2024}%
  \BibitemOpen
  \bibfield  {author} {\bibinfo {author} {\bibfnamefont {L.}~\bibnamefont
  {Fritz}}\ and\ \bibinfo {author} {\bibfnamefont {T.}~\bibnamefont
  {Scaffidi}},\ }\bibfield  {title} {\bibinfo {title} {Hydrodynamic electronic
  transport},\ }\href@noop {} {\bibfield  {journal} {\bibinfo  {journal} {Annu.
  Rev. Condens. Matter Phys.}\ }\textbf {\bibinfo {volume} {15}},\ \bibinfo
  {pages} {17} (\bibinfo {year} {2024})}\BibitemShut {NoStop}%
\bibitem [{\citenamefont {Wu}\ \emph {et~al.}(2022)\citenamefont {Wu},
  \citenamefont {Liao},\ and\ \citenamefont {Foster}}]{WuLiaoFoster2022}%
  \BibitemOpen
  \bibfield  {author} {\bibinfo {author} {\bibfnamefont {T.~C.}\ \bibnamefont
  {Wu}}, \bibinfo {author} {\bibfnamefont {Y.}~\bibnamefont {Liao}},\ and\
  \bibinfo {author} {\bibfnamefont {M.~S.}\ \bibnamefont {Foster}},\ }\bibfield
   {title} {\bibinfo {title} {Quantum interference of hydrodynamic modes in a
  dirty marginal {F}ermi liquid},\ }\href@noop {} {\bibfield  {journal}
  {\bibinfo  {journal} {Phys. Rev. B}\ }\textbf {\bibinfo {volume} {106}},\
  \bibinfo {pages} {155108} (\bibinfo {year} {2022})}\BibitemShut {NoStop}%
\bibitem [{\citenamefont {Hartnoll}\ \emph {et~al.}(2018)\citenamefont
  {Hartnoll}, \citenamefont {Lucas},\ and\ \citenamefont {Sachdev}}]{HLS2018}%
  \BibitemOpen
  \bibfield  {author} {\bibinfo {author} {\bibfnamefont {S.~A.}\ \bibnamefont
  {Hartnoll}}, \bibinfo {author} {\bibfnamefont {A.}~\bibnamefont {Lucas}},\
  and\ \bibinfo {author} {\bibfnamefont {S.}~\bibnamefont {Sachdev}},\
  }\href@noop {} {\emph {\bibinfo {title} {Holographic Quantum Matter}}}\
  (\bibinfo  {publisher} {MIT Press},\ \bibinfo {address} {Cambridge, MA},\
  \bibinfo {year} {2018})\BibitemShut {NoStop}%
\bibitem [{\citenamefont {Grozdanov}\ \emph {et~al.}(2019)\citenamefont
  {Grozdanov}, \citenamefont {Lucas},\ and\ \citenamefont
  {Poovuttikul}}]{GLP2019}%
  \BibitemOpen
  \bibfield  {author} {\bibinfo {author} {\bibfnamefont {S.}~\bibnamefont
  {Grozdanov}}, \bibinfo {author} {\bibfnamefont {A.}~\bibnamefont {Lucas}},\
  and\ \bibinfo {author} {\bibfnamefont {N.}~\bibnamefont {Poovuttikul}},\
  }\bibfield  {title} {\bibinfo {title} {Holography and hydrodynamics with
  weakly broken symmetries},\ }\href@noop {} {\bibfield  {journal} {\bibinfo
  {journal} {Phys. Rev. D}\ }\textbf {\bibinfo {volume} {99}},\ \bibinfo
  {pages} {086012} (\bibinfo {year} {2019})}\BibitemShut {NoStop}%
\bibitem [{\citenamefont {Gell-Mann}\ \emph {et~al.}(1968)\citenamefont
  {Gell-Mann}, \citenamefont {Oakes},\ and\ \citenamefont {Renner}}]{GMOR1968}%
  \BibitemOpen
  \bibfield  {author} {\bibinfo {author} {\bibfnamefont {M.}~\bibnamefont
  {Gell-Mann}}, \bibinfo {author} {\bibfnamefont {R.~J.}\ \bibnamefont
  {Oakes}},\ and\ \bibinfo {author} {\bibfnamefont {B.}~\bibnamefont
  {Renner}},\ }\bibfield  {title} {\bibinfo {title} {Behavior of current
  divergences under {SU(3)}$\times${SU(3)}},\ }\href@noop {} {\bibfield
  {journal} {\bibinfo  {journal} {Phys. Rev.}\ }\textbf {\bibinfo {volume}
  {175}},\ \bibinfo {pages} {2195} (\bibinfo {year} {1968})}\BibitemShut
  {NoStop}%
\bibitem [{\citenamefont {Gasser}\ and\ \citenamefont
  {Leutwyler}(1984)}]{GasserLeutwyler1984}%
  \BibitemOpen
  \bibfield  {author} {\bibinfo {author} {\bibfnamefont {J.}~\bibnamefont
  {Gasser}}\ and\ \bibinfo {author} {\bibfnamefont {H.}~\bibnamefont
  {Leutwyler}},\ }\bibfield  {title} {\bibinfo {title} {Chiral perturbation
  theory to one loop},\ }\href@noop {} {\bibfield  {journal} {\bibinfo
  {journal} {Ann. Phys.}\ }\textbf {\bibinfo {volume} {158}},\ \bibinfo {pages}
  {142} (\bibinfo {year} {1984})}\BibitemShut {NoStop}%
\bibitem [{\citenamefont {Andreev}\ \emph {et~al.}(2011)\citenamefont
  {Andreev}, \citenamefont {Kivelson},\ and\ \citenamefont {Spivak}}]{AKS2011}%
  \BibitemOpen
  \bibfield  {author} {\bibinfo {author} {\bibfnamefont {A.~V.}\ \bibnamefont
  {Andreev}}, \bibinfo {author} {\bibfnamefont {S.~A.}\ \bibnamefont
  {Kivelson}},\ and\ \bibinfo {author} {\bibfnamefont {B.}~\bibnamefont
  {Spivak}},\ }\bibfield  {title} {\bibinfo {title} {Hydrodynamic description
  of transport in strongly correlated electron systems},\ }\href@noop {}
  {\bibfield  {journal} {\bibinfo  {journal} {Phys. Rev. Lett.}\ }\textbf
  {\bibinfo {volume} {106}},\ \bibinfo {pages} {256804} (\bibinfo {year}
  {2011})}\BibitemShut {NoStop}%
\bibitem [{\citenamefont {Torquato}(2002)}]{Torquato2002}%
  \BibitemOpen
  \bibfield  {author} {\bibinfo {author} {\bibfnamefont {S.}~\bibnamefont
  {Torquato}},\ }\href@noop {} {\emph {\bibinfo {title} {Random Heterogeneous
  Materials}}}\ (\bibinfo  {publisher} {Springer},\ \bibinfo {address} {New
  York},\ \bibinfo {year} {2002})\BibitemShut {NoStop}%
\bibitem [{\citenamefont {Brinkman}(1949)}]{Brinkman1949}%
  \BibitemOpen
  \bibfield  {author} {\bibinfo {author} {\bibfnamefont {H.~C.}\ \bibnamefont
  {Brinkman}},\ }\bibfield  {title} {\bibinfo {title} {A calculation of the
  viscous force exerted by a flowing fluid on a dense swarm of particles},\
  }\href@noop {} {\bibfield  {journal} {\bibinfo  {journal} {Appl. Sci. Res.
  A}\ }\textbf {\bibinfo {volume} {1}},\ \bibinfo {pages} {27} (\bibinfo {year}
  {1949})}\BibitemShut {NoStop}%
\bibitem [{\citenamefont {Kim}\ and\ \citenamefont
  {Russel}(1985)}]{KimRussel1985}%
  \BibitemOpen
  \bibfield  {author} {\bibinfo {author} {\bibfnamefont {S.}~\bibnamefont
  {Kim}}\ and\ \bibinfo {author} {\bibfnamefont {W.~B.}\ \bibnamefont
  {Russel}},\ }\bibfield  {title} {\bibinfo {title} {Modelling of porous media
  by renormalization of the {S}tokes equations},\ }\href@noop {} {\bibfield
  {journal} {\bibinfo  {journal} {J. Fluid Mech.}\ }\textbf {\bibinfo {volume}
  {154}},\ \bibinfo {pages} {269} (\bibinfo {year} {1985})}\BibitemShut
  {NoStop}%
\bibitem [{\citenamefont {Paw{\l}owski}\ \emph {et~al.}(2026)\citenamefont
  {Paw{\l}owski}, \citenamefont {Sur{\'o}wka},\ and\ \citenamefont
  {Zarembo}}]{Pawlowski2026}%
  \BibitemOpen
  \bibfield  {author} {\bibinfo {author} {\bibfnamefont {J.}~\bibnamefont
  {Paw{\l}owski}}, \bibinfo {author} {\bibfnamefont {P.}~\bibnamefont
  {Sur{\'o}wka}},\ and\ \bibinfo {author} {\bibfnamefont {K.}~\bibnamefont
  {Zarembo}},\ }\href@noop {} {\bibinfo {title} {Geometry-induced skin effect
  in electron hydrodynamics}} (\bibinfo {year} {2026}),\ \bibinfo {note}
  {arXiv:2601.17229}\BibitemShut {NoStop}%
\bibitem [{\citenamefont {Dykhne}(1971)}]{Dykhne1971}%
  \BibitemOpen
  \bibfield  {author} {\bibinfo {author} {\bibfnamefont {A.~M.}\ \bibnamefont
  {Dykhne}},\ }\bibfield  {title} {\bibinfo {title} {Conductivity of a
  two-dimensional two-phase system},\ }\href@noop {} {\bibfield  {journal}
  {\bibinfo  {journal} {Sov. Phys. JETP}\ }\textbf {\bibinfo {volume} {32}},\
  \bibinfo {pages} {63} (\bibinfo {year} {1971})}\BibitemShut {NoStop}%
\bibitem [{\citenamefont {Matheron}(1967)}]{Matheron1967}%
  \BibitemOpen
  \bibfield  {author} {\bibinfo {author} {\bibfnamefont {G.}~\bibnamefont
  {Matheron}},\ }\href@noop {} {\emph {\bibinfo {title} {{\'E}l{\'e}ments pour
  une th{\'e}orie des milieux poreux}}}\ (\bibinfo  {publisher} {Masson},\
  \bibinfo {address} {Paris},\ \bibinfo {year} {1967})\BibitemShut {NoStop}%
\bibitem [{\citenamefont {Lucas}(2017)}]{Lucas2017Stokes}%
  \BibitemOpen
  \bibfield  {author} {\bibinfo {author} {\bibfnamefont {A.}~\bibnamefont
  {Lucas}},\ }\bibfield  {title} {\bibinfo {title} {Stokes paradox in
  electronic {F}ermi liquids},\ }\href@noop {} {\bibfield  {journal} {\bibinfo
  {journal} {Phys. Rev. B}\ }\textbf {\bibinfo {volume} {95}},\ \bibinfo
  {pages} {115425} (\bibinfo {year} {2017})}\BibitemShut {NoStop}%
\bibitem [{\citenamefont {Mirlin}\ \emph {et~al.}(2001)\citenamefont {Mirlin},
  \citenamefont {Polyakov}, \citenamefont {Evers},\ and\ \citenamefont
  {W\"olfle}}]{Mirlin2001}%
  \BibitemOpen
  \bibfield  {author} {\bibinfo {author} {\bibfnamefont {A.~D.}\ \bibnamefont
  {Mirlin}}, \bibinfo {author} {\bibfnamefont {D.~G.}\ \bibnamefont
  {Polyakov}}, \bibinfo {author} {\bibfnamefont {F.}~\bibnamefont {Evers}},\
  and\ \bibinfo {author} {\bibfnamefont {P.}~\bibnamefont {W\"olfle}},\
  }\bibfield  {title} {\bibinfo {title} {Quasiclassical negative
  magnetoresistance of a {2D} electron gas: interplay of strong scatterers and
  smooth disorder},\ }\href@noop {} {\bibfield  {journal} {\bibinfo  {journal}
  {Phys. Rev. Lett.}\ }\textbf {\bibinfo {volume} {87}},\ \bibinfo {pages}
  {126805} (\bibinfo {year} {2001})}\BibitemShut {NoStop}%
\bibitem [{\citenamefont {Raskulov}\ \emph {et~al.}(2026)\citenamefont
  {Raskulov}, \citenamefont {Baryshnikov},\ and\ \citenamefont
  {Alekseev}}]{Raskulov2026}%
  \BibitemOpen
  \bibfield  {author} {\bibinfo {author} {\bibfnamefont {D.~R.}\ \bibnamefont
  {Raskulov}}, \bibinfo {author} {\bibfnamefont {K.~A.}\ \bibnamefont
  {Baryshnikov}},\ and\ \bibinfo {author} {\bibfnamefont {P.~S.}\ \bibnamefont
  {Alekseev}},\ }\href@noop {} {\bibinfo {title} {Hydrodynamics of
  two-dimensional electrons due to scattering by disorder}} (\bibinfo {year}
  {2026}),\ \bibinfo {note} {arXiv:2609.04097}\BibitemShut {NoStop}%
\bibitem [{\citenamefont {Forster}\ \emph {et~al.}(1977)\citenamefont
  {Forster}, \citenamefont {Nelson},\ and\ \citenamefont {Stephen}}]{FNS1977}%
  \BibitemOpen
  \bibfield  {author} {\bibinfo {author} {\bibfnamefont {D.}~\bibnamefont
  {Forster}}, \bibinfo {author} {\bibfnamefont {D.~R.}\ \bibnamefont
  {Nelson}},\ and\ \bibinfo {author} {\bibfnamefont {M.~J.}\ \bibnamefont
  {Stephen}},\ }\bibfield  {title} {\bibinfo {title} {Large-distance and
  long-time properties of a randomly stirred fluid},\ }\href@noop {} {\bibfield
   {journal} {\bibinfo  {journal} {Phys. Rev. A}\ }\textbf {\bibinfo {volume}
  {16}},\ \bibinfo {pages} {732} (\bibinfo {year} {1977})}\BibitemShut
  {NoStop}%
\bibitem [{\citenamefont {Kovtun}\ \emph {et~al.}(2011)\citenamefont {Kovtun},
  \citenamefont {Moore},\ and\ \citenamefont {Romatschke}}]{KMR2011}%
  \BibitemOpen
  \bibfield  {author} {\bibinfo {author} {\bibfnamefont {P.}~\bibnamefont
  {Kovtun}}, \bibinfo {author} {\bibfnamefont {G.~D.}\ \bibnamefont {Moore}},\
  and\ \bibinfo {author} {\bibfnamefont {P.}~\bibnamefont {Romatschke}},\
  }\bibfield  {title} {\bibinfo {title} {Stickiness of sound: An absolute lower
  limit on viscosity and the breakdown of second-order relativistic
  hydrodynamics},\ }\href@noop {} {\bibfield  {journal} {\bibinfo  {journal}
  {Phys. Rev. D}\ }\textbf {\bibinfo {volume} {84}},\ \bibinfo {pages} {025006}
  (\bibinfo {year} {2011})}\BibitemShut {NoStop}%
\bibitem [{\citenamefont {Keller}(1964)}]{Keller1964}%
  \BibitemOpen
  \bibfield  {author} {\bibinfo {author} {\bibfnamefont {J.~B.}\ \bibnamefont
  {Keller}},\ }\bibfield  {title} {\bibinfo {title} {A theorem on the
  conductivity of a composite medium},\ }\href@noop {} {\bibfield  {journal}
  {\bibinfo  {journal} {J. Math. Phys.}\ }\textbf {\bibinfo {volume} {5}},\
  \bibinfo {pages} {548} (\bibinfo {year} {1964})}\BibitemShut {NoStop}%
\bibitem [{\citenamefont {Kato}(1995)}]{Kato1995}%
  \BibitemOpen
  \bibfield  {author} {\bibinfo {author} {\bibfnamefont {T.}~\bibnamefont
  {Kato}},\ }\href@noop {} {\emph {\bibinfo {title} {Perturbation Theory for
  Linear Operators}}}\ (\bibinfo  {publisher} {Springer},\ \bibinfo {address}
  {Berlin},\ \bibinfo {year} {1995})\BibitemShut {NoStop}%
\end{thebibliography}
%

\end{document}